\documentclass[12pt]{article}

\input{epsf}
\usepackage[nosort]{cite}
\usepackage{graphicx}
\usepackage{color}
\usepackage{amsmath}
\usepackage{epsfig}
\usepackage{amsfonts}
\usepackage{amssymb}
\usepackage{array}

\newcommand{\be}{\begin{equation}}
\newcommand{\ee}{\end{equation}}
\newcommand{\bee}{\begin{equation*}}
\newcommand{\eee}{\end{equation*}}
\newcommand{\bea}{\begin{eqnarray}}
\newcommand{\eea}{\end{eqnarray}}
\newcommand{\bean}{\begin{eqnarray*}}
\newcommand{\eean}{\end{eqnarray*}}

\begin{document}

\setcounter{page}{0}
\thispagestyle{empty}

\begin{flushright}
ULB-TH/12-18
LPT 12-113
\end{flushright}

\vskip 8pt

\begin{center}
{\bf \LARGE {Non-thermal Dark Matter Production \\
\vskip 8pt
from the Electroweak Phase Transition:}}
\end{center}

\vspace{-6mm}

\begin{center}
{\bf \LARGE {Multi-TeV WIMPs and ``Baby-Zillas"}}
\end{center}

\vskip 12pt

\begin{center}
{\bf   Adam Falkowski$^{a}$ and
Jose M. No$^{b,c}$}
\end{center}

\vskip 20pt

\begin{center}

\centerline{$^{a}${\it Laboratoire de Physique Th\'eorique d'Orsay, UMR8627-CNRS}}
\centerline{{\it Universit\'e Paris-Sud, 91405 Orsay, France}}
\centerline{$^{b}${\it Service de Physique Th\'eorique, Universit\'e Libre de Bruxelles, 
B-1050 Bruxelles, Belgium}}
\centerline{$^{c}${\it University of Sussex, Department of Physics and Astronomy}}
\centerline{{\it BN1 9QH Brighton, United Kingdom}}
\vskip .3cm
\centerline{\tt adam.falkowski@th.u-psud.fr, J.M.No@sussex.ac.uk}
\end{center}

\vskip 13pt

\begin{abstract}
Particle production at the end of a first-order electroweak phase transition may be rather generic in theories 
beyond the standard model. 
Dark matter may then be abundantly produced by this mechanism if it has a sizable coupling to the Higgs field. 
For an electroweak phase transition occuring at a temperature
$T_{\mathrm{EW}} \sim 50-100$ GeV, non-thermally generated dark matter with mass $M_X >$ TeV 
will survive thermalization after the phase transition, and could then potentially account for 
the observed dark matter relic density in scenarios where a thermal dark matter component is 
either too small or absent. 
Dark matter in these scenarios could then either be multi-TeV WIMPs whose relic abundace is mostly generated 
at the electroweak phase transition, or \textit{``Baby-Zillas"} with mass 
$M_{\mathrm{GUT}} \gg M_X \gg v_{\mathrm{EW}}$  that never reach  thermal equilibrium in the early universe. 
\end{abstract}

\newpage
\renewcommand{\theequation}{\arabic{section}.\arabic{equation}}

\section{Introduction}

The most popular paradigm for the origin of dark matter (DM) in the Universe is the {\em thermal freeze-out}. 
In that scenario, the dark matter particle with mass $M_X$ annihilates into matter with a cross section 
$\left\langle \sigma \, v \right\rangle_{\rm thermal} \sim 3 \times 10^{-26} {\rm  cm}^3/{\rm s}$.
This ensures dark matter is in thermal equilibrium  with the rest of the plasma in the early universe 
while $T \gtrsim M$ but decouples when $T \sim M_X/20$, leaving the relic abundance in agreement with 
the value $\Omega_{X} = 0.228 \pm 0.027$ measured by WMAP \cite{Komatsu:2010fb}.  
Incidentally,  $\left\langle \sigma \, v \right\rangle_{\rm thermal}$ is a generic cross section for a 
weak scale mass  
particle interacting with order one couplings, this fact being referred to as the {\em WIMP miracle}. 
In spite of these attractive features, 
non-thermal mechanisms of dark matter production have also received considerable attention. 
Examples include right-handed neutrinos produced by oscillations \cite{Dodelson:1993je}, 
axions produced by vacuum misalignment \cite{Preskill:1982cy},  winos produced from moduli 
decays \cite{Moroi:1999zb}, and super-massive dark matter ({\em WIMP-zillas}) produced 
during reheating after inflation \cite{Wimpzillas}. These studies allow one to recognize a 
wider range of possible collider and astrophysical signals of dark matter than what would result from
the thermal WIMP scenario. 

In this paper we study the possibility of non-thermal dark matter production during a 
first-order electroweak (EW) phase transition. Bubble collisions at the end of the EW phase 
transition may give rise to abundant non-thermal particle production when a sizable amount 
of the energy budget of the transition is stored in the bubble walls, possibly leading to 
new and appealing scenarios.  Many models of dark matter contain a direct coupling between the Higgs 
and the dark matter candidate fields (MSSM and its extensions, Little Higgs theories 
with T-parity and DM extensions of the  standard model (SM) via the Higgs portal, to name a few). 
It is thus reasonable to expect that dark matter may be abundantly produced non-thermally at the end of 
a first-order EW phase transition. Note that, much like in the thermal WIMP case, dark matter would then be a particle 
with $M_X \sim 10 \, \mathrm{GeV} - 10 \, \mathrm{TeV}$ with significant coupling to the SM, thus being  
within reach of colliders and DM direct detection experiments.
 
There is however one generic problem with this scenario. Since the temperature of the Universe right after the EW phase transition is 
$T_{\mathrm{EW}} \sim 50-100$ GeV (for strong transitions $T_{\mathrm{EW}}$ may be somewhat lower 
than $100$ GeV),  thermalization will typically lead to a wash-out of the non-thermal abundance, thus rendering the particle production 
at the EW phase transition irrelevant for  the subsequent evolution of the Universe.
The wash-out process can nevertheless be avoided in a number of ways,
each resulting in a scenario where non-thermal dark matter production is (fully or partially) responsible for the 
observed dark matter relic density.  One possibility, recently outlined in \cite{Konstandin:2011dr}, is to allow for a few e-foldings 
of inflation prior to the beginning of the transition (which can happen for very strong EW phase transitions), 
diluting the plasma and leaving the Universe partially empty. If the reheating temperature 
after the phase transition is low, $T_{\mathrm{RH}} \ll 100$ GeV, it may be possible for a dark 
matter candidate with weak couplings to the Higgs field and mass $M_X \sim 100$ GeV to remain 
out of thermal equilibrium after the EW phase transition.  
In this paper we investigate other scenarios allowing for a survival of the non-thermal abundance.   

One possibility corresponds to the case of relatively heavy (multi-TeV) dark matter: for $M_X \gtrsim 1$~TeV, dark matter 
will be very non-relativistic when the EW phase transition takes place, and the decoupling/freeze-out 
temperature $T_{\mathrm{FO}}$ will satisfy $T_{\mathrm{FO}} \sim M_X / 20 \gtrsim T_{\mathrm{EW}}$. 
Then, heavy dark matter produced non-thermally through bubble collisions may remain out of thermal equilibrium 
after the EW phase transition (or at least wash-out will be partially avoided). 
Another possibility is that bubble collisions produce super-heavy dark matter,  $M_X \sim 10^6$-$10^8$ GeV, 
a scenario we call \textit{``baby-zillas"}.  
We argue this may be possible for a very strong EW phase transition and dark matter having a large coupling the Higgs. 
In order for baby-zillas with $M_X \gg v_{\mathrm{EW}}$ to  be a viable dark matter candidate, 
they must have never reached thermal equilibrium in the early universe after inflation, since 
otherwise they would have over-closed the universe.  
This sets a relatively low upper bound on the reheating  temperature after inflation in that scenario. 
Finally, asymmetric dark matter production might allow to avoid complete wash-out
of the non-thermal abundance through thermalization after the EW phase transition.

The paper is organized as follows: in Section~2 we review the formalism 
that describes particle production at the end of the EW phase transition for the case 
of very elastic bubble collisions \cite{Hawking:1982ga,Watkins:1991zt} and extend it to the 
case of very inelastic ones, highlighting the differences between 
both scenarios \cite{KS}. Then, in Section~3 we explicitly 
compute the particle production efficiency of scalar, fermion, and vector boson particles 
coupled to the 
Higgs (either directly or via an effective Higgs portal).
In Sections~4~and~5 we focus on dark matter production at the end of the EW phase transition.
 First we discuss in Section~4 the conditions for non-thermally 
produced dark matter to avoid subsequent 
wash-out and constitute the bulk of the present dark matter density, selecting heavy 
(multi-TeV) vector boson dark matter as a viable example.  
We go on to analyze in detail non-thermal dark matter production in that scenario and the subsequent evolution
of the non-thermally generated abundance after the EW phase transition, including finally 
a discussion on the current XENON100 bounds and direct detection prospects.
Then, in Section~5 we study the non-thermal production of very heavy ($M_X \gg v_{\mathrm{EW}}$) vector boson 
dark matter, and outline the conditions under which these {\em baby-zillas} constitute a viable dark matter 
candidate. 
In the case of asymmetric non-thermal dark  matter production, we find it difficult to avoid subsequent wash-out, and the discussion is left for an appendix.  
We summarize our results and conclude in Section 6.

\section{Particle Production at the EW Phase Transition}

\subsection{Bubble Collisions in the EW Phase Transition}
\label{section21}

If the early Universe was hotter than $T_{\rm EW} \sim 100$ GeV it must have undergone an EW 
phase transition at some point in its history. Within the SM, the EW phase transition is a smooth cross-over, 
however it is conceivable that new degrees of freedom beyond the SM modify the Higgs potential so as to 
make the transition first order. This is what we assume throughout this paper, without specifying the 
full theory that makes the first order transition possible. In that case, the EW phase transition 
proceeded through nucleation and expansion of bubbles of true Higgs vacuum, which eventually 
collided among each other completing the transition. As this was happening during the
radiation dominated era, the bubble expansion process would then have taken place in a thermal 
environment (except for the case when a period of inflation would have preceded the phase transition).  

For a first order phase transition occuring in a thermal environment, the study of the bubble expansion 
process reveals that the thermal plasma exerts some amount of friction on the expanding bubble wall, 
and this friction tends to balance the pressure difference on the bubble wall driving the expansion. 
In the usual picture, nucleated bubbles reach 
a stationary state after a very short period of acceleration, with a constant wall velocity 
depending specifically on the interactions of the bubble wall with the degrees of freedom in the 
plasma \cite{Friction1,Friction2} and on the resulting 
fluid dynamics \cite{Steinhardt:1981ct,Gyulassy:1983rq,Laine:1993ey} (see \cite{EKNS} for a review). 
In this case, the amount of energy stored in the bubble walls at the time of the bubble collisions
is negligible compared to the available energy of the transition, since most of this available energy gets 
converted into plasma bulk motion and thermal energy \cite{Kamionkowski:1993fg}. 

However, this picture was recently challenged in \cite{BM}, where it was shown that the friction 
exerted by the plasma may saturate to a finite value for ultrarelativistic bubble walls. As a consequence  
the stationary state assumption will no longer be true when the pressure difference on the bubble wall is larger than the friction
saturation value, which may happen for strongly first order phase transitions. In this scenario, if there
are no hydrodynamic obstacles that prohibit the bubble walls 
to become highly relativistic in the first place (see however \cite{KN}), bubbles will expand in an accelerated way
(`the so-called {\em runaway  bubbles}), with almost all the energy of the transition being used to accelerate the bubble 
walls\footnote[1]{This situation may also arise if, under very specific circumstances, a few e-foldings of inflation 
are achieved prior to the beginning of the EW phase transition (see \cite{Konstandin:2011dr} for a natural realization 
of this scenario), diluting the plasma 
and leaving the Universe mostly empty. In this case the expansion of the bubbles effectively takes 
place in vacuum, and the nucleated bubbles expand in an accelerated way due to the absence of friction.}\cite{EKNS}. 
By the end of the phase transition (when bubbles start colliding), these runaway bubbles may reach very large values of $\gamma_w$:

\begin{equation}
\label{eq:gammaw}
\gamma_w \lesssim \gamma^{\mathrm{max}}_w \sim \frac{\beta^{-1}}{H^{-1}}\, 
\frac{M_{\mathrm{pl}}}{v_T} \sim 10^{15}\, ,
\end{equation}

\noindent with $v_{T}$ the value of the Higgs VEV in the broken phase and 
$\beta^{-1} \sim \left( 10^{-3} - 10^{-2} \right) H^{-1}$ being the duration of the phase 
transition \cite{Hogan}. 
The estimate (\ref{eq:gammaw}) follows from balancing the surface energy on the bubble wall 
(\ref{energywall}) and the energy available inside the bubble. 

Once bubbles start colliding, the energy stored on the bubble walls will be liberated into the plasma.
As argued above, for ``runaway" bubbles this will correspond to a very large portion of the energy 
budget of the phase transition, and therefore this process can be very important. Under certain circumstances,
this may also hold true for highly relativistic bubble walls ($\gamma_w \gg 1$) that reach
a stationary state long before bubble collisions start (meaning that $\gamma_w \ll \gamma^{\mathrm{max}}_w$), 
in which case the amount of energy stored in the bubble walls will be very small compared to the available energy of the transition,
but still important when released into the plasma at the end of the transition.

The process of bubble collisions 
in cosmological first order phase transitions is by itself a very complicated one. 
Consider a configuration of two planar bubble walls\footnote[2]{At the time of the collision, the 
bubbles are so large compared to the relevant microscopical scales, that their walls 
may be considered planar as a good approximation.} initially far away from each other, that approach and 
collide \cite{Hawking:1982ga,Watkins:1991zt,Chinos}. 
Depending of the shape of the potential for the scalar field $\phi$ driving the transition (in our case, the Higgs field $h$), 
the bubble collision will be approximately elastic or partially inelastic \cite{Hawking:1982ga,Watkins:1991zt} 
(see also \cite{KS}). In the first case, the walls reflect off one another after the collision, which reestablishes a region
of symmetric phase between the bubble walls. For a perfectly elastic collision the field profile of the colliding walls 
in the limit of infinitely thin bubble walls (taken as step-functions) can be written as \cite{Watkins:1991zt}

\be
\label{2WallsFieldConfiguration}
h(z,t) = h_{\infty} \equiv \left\lbrace\begin{array}{lll}
0 & \mathrm{if} \, \, v_w\, t < z < -v_w\, t & \quad t < 0, \\ 
0 & \mathrm{if} \, -v_w\, t < z < v_w\, t & \quad t > 0, \\
v_{T} & \mathrm{Otherwise},
\end{array} \right.
\ee

\noindent where $v_w$ is the bubble wall velocity, 
the bubble walls move in the $z$-direction and the collision is assumed to happen at $t = 0$. 
Since we are ultimately interested in scenarios where 
$\gamma_w \gg 1$, we will take the ultrarelativistic limit $v_w \rightarrow 1$ in the rest of the section. 
The field profile (\ref{2WallsFieldConfiguration}) neglects the thickness of the bubble walls $l_w$ 
(generically, $l_w \sim (10-30)/T_{\mathrm{EW}}$, with 
$T_{\mathrm{EW}} \sim 50 - 100$ GeV). 
To capture the wall thickness effects one can consider another ansatz for the profile of the colliding 
bubble walls: 

\bea
\label{2WallsFieldConfiguration2}
h(z,t) = h_{l_{w}} \equiv \frac{v_{T}}{2} \left[\mathrm{Tanh}\left(\gamma_w\frac{t + \left| z \right|}{l_w}\right) - 
\mathrm{Tanh}\left(\gamma_w\frac{t - \left| z \right|}{l_w}\right) \right] \nonumber \\
= \frac{v_{T}}{2} \left[2 + \mathrm{Tanh}\left(\gamma_w\frac{z -\left| t \right|}{l_w}\right) - 
\mathrm{Tanh}\left(\gamma_w\frac{z + \left| t \right|}{l_w}\right) \right].
\eea 

A perfectly elastic collision is 
however an idealized situation, as one expects a certain degree of inelasticity in a realistic collision.
Moreover, even for a very elastic collision the bubble walls will eventually be drawn back together by 
vacuum pressure and collide 
again. A quantitative picture of the collision of two planar bubble walls can be obtained by studying the evolution 
equation for the scalar field configuration $h(z,t)$ subject to the potential $V(h)$:

\be
\label{eqmotion}
\left(\partial^2_t  - \partial^2_z \right) h(z,t) = -\frac{\partial V(h)}{\partial h},
\ee

\noindent with the initial condition corresponding to two planar bubble walls far away from 
each other and moving in opposite directions (given approximately by $h_{l_{w}}$ in the 
limit $t \rightarrow - \infty$). 
In the ultrarelativistic limit the ansatz (\ref{2WallsFieldConfiguration2}) 
will also be an approximate solution of (\ref{eqmotion}) before the bubble collision\footnote[3]{
Each bubble wall in (\ref{2WallsFieldConfiguration2}) interpolates between the symmetric
and broken minima of $V(h)$, and so $\partial V(h)/\partial h = 0$ outside the bubble wall. Then, 
for very thin walls the 
equation of motion approximately simplifies (before the collision) to 
$\left(\partial^2_t  - \partial^2_z \right) h(z,t) = 0$, 
for which any function of the form $f(z+t)$ or $f(z-t)$ is a solution.} (for $t < 0$). In 
this limit, the kinetic energy per unit 
area contained in the field configuration $h(z,t)$ prior to the collision is given by 

\be
\label{energywall}
\frac{E_w}{A} = \frac{2}{3}\, v_{T}^2 \,\frac{\gamma_w}{l_w}.
\ee 

\begin{figure}[ht]
\begin{center}
\includegraphics[width=0.45\textwidth, clip ]{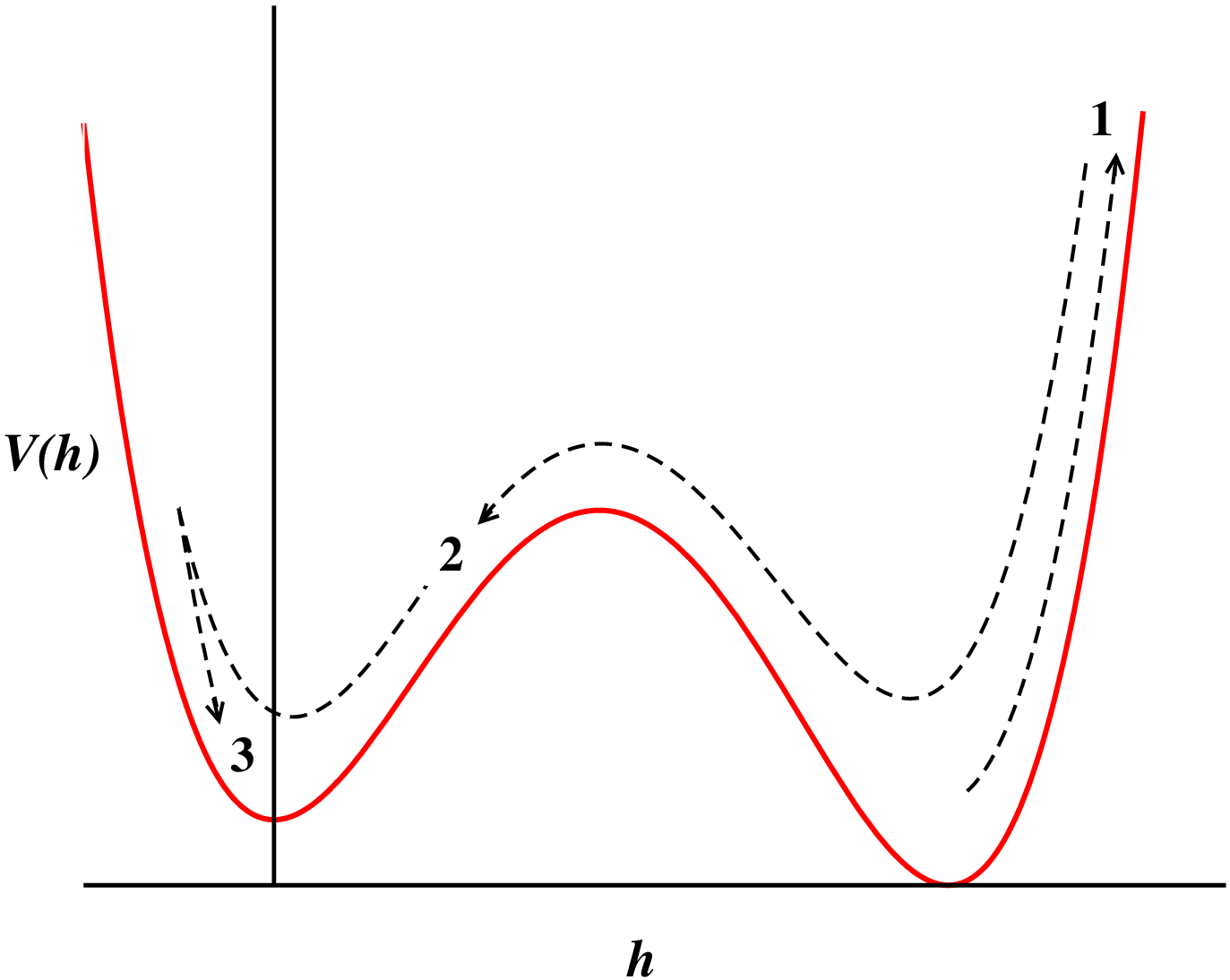} \hspace{5mm}
\includegraphics[width=0.45\textwidth, clip ]{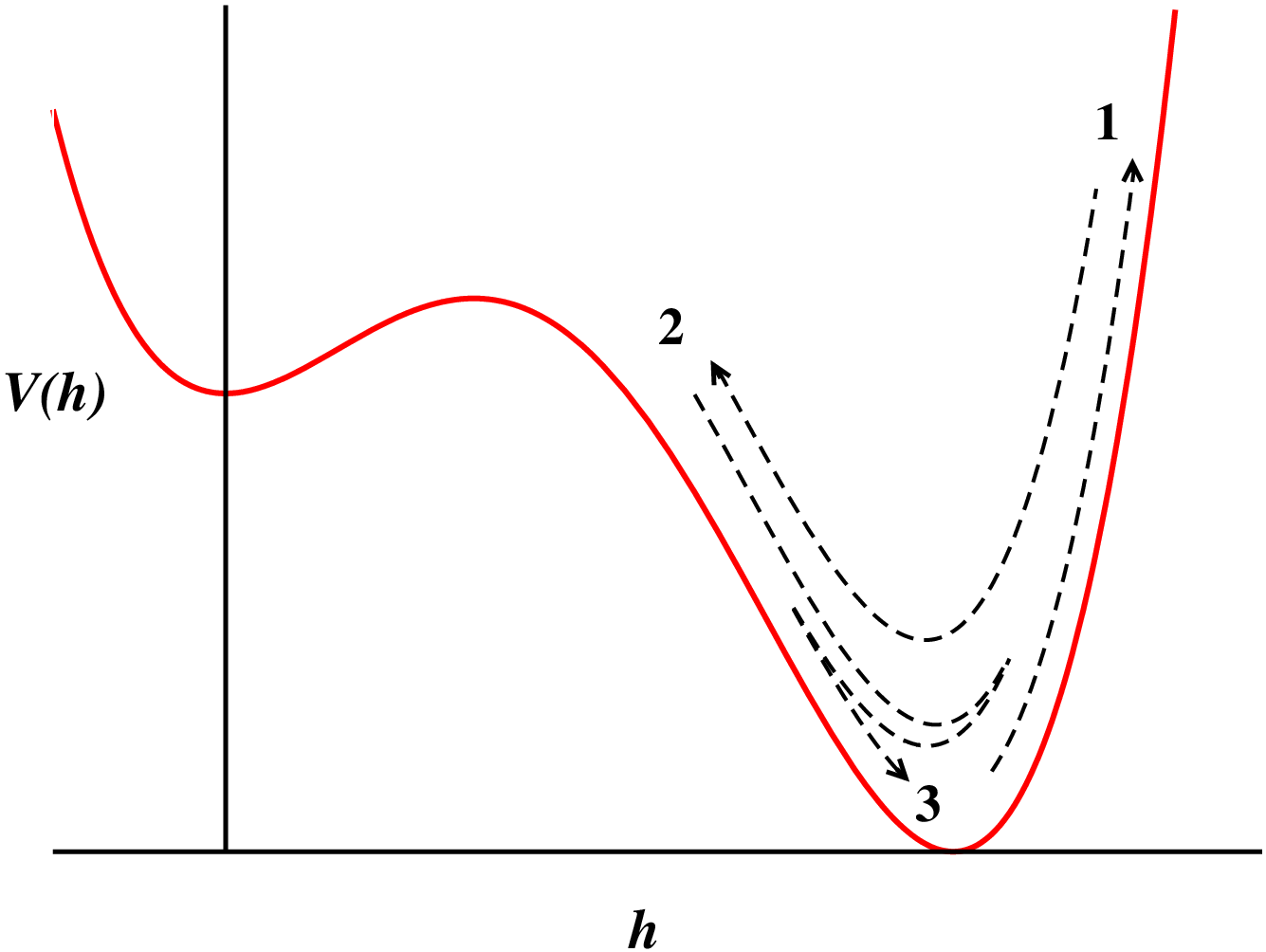} 
\caption{\small LEFT: Potential with nearly degenerate minima. RIGHT: Potential with very non-degenerate minima. 
Each one shows the behaviour of the field immediately after the collision in the region close to the collision 
point, as described in the text: 1) ``Kick" to field values larger than $v(T)$. 2) Large field oscillation, successful
(LEFT) or unsuccessful (RIGHT) in driving the field over the potential barrier. 3) Oscillations 
around the symmetric (LEFT) or broken (RIGHT) minimum.}
\label{Fig:3}
\end{center}
\end{figure}

At the moment of the collision, the field configuration makes an ``excursion" to field values larger 
than $v_{T}$ in a small region around the collision 
point \cite{Chinos} (resulting in $\partial V /\partial h \neq 0$ in this region). 
The subsequent evolution 
of $h(z,t)$ strongly depends on the shape of the potential $V(h)$. 
The field close to the collision region oscillates back after the initial ``kick" in field space, and for a potential 
with nearly degenerate minima this oscillation is able to drive the field over the potential barrier and 
into the basin of attraction of the symmetric minimum (Figure \ref{Fig:3} - Left), where it will perform small-amplitude 
oscillations. In this case the collision is approximately elastic as described above, with 
the bubble walls being effectively reflected as a region of symmetric phase is re-established between them. 
The walls move then away from each other until vacuum pressure makes them approach and collide again, repeating the 
process several times. In each collision some fraction of the energy stored in the walls is radiated 
into scalar waves and quanta of the fields coupled to $h$, until all of the energy in the walls is radiated away.  
In contrast to this scenario, for a potential $V(h)$ 
with very non-degenerate minima (Figure \ref{Fig:3} - Right), the field oscillation after the ``kick" in the region 
close to the collision point does not effectively drive the field over the potential barrier. As a consequence, the field 
stays in the basin of attraction of the broken minimum $v_T$ and performs relatively large-amplitude oscillations around it,
giving rise to a large amount of energy radiated into scalar waves (as opposed to the previous scenario). 
In this case the collision is very inelastic. 

\begin{figure}[ht]
\begin{center}
\includegraphics[width=0.49\textwidth, clip ]{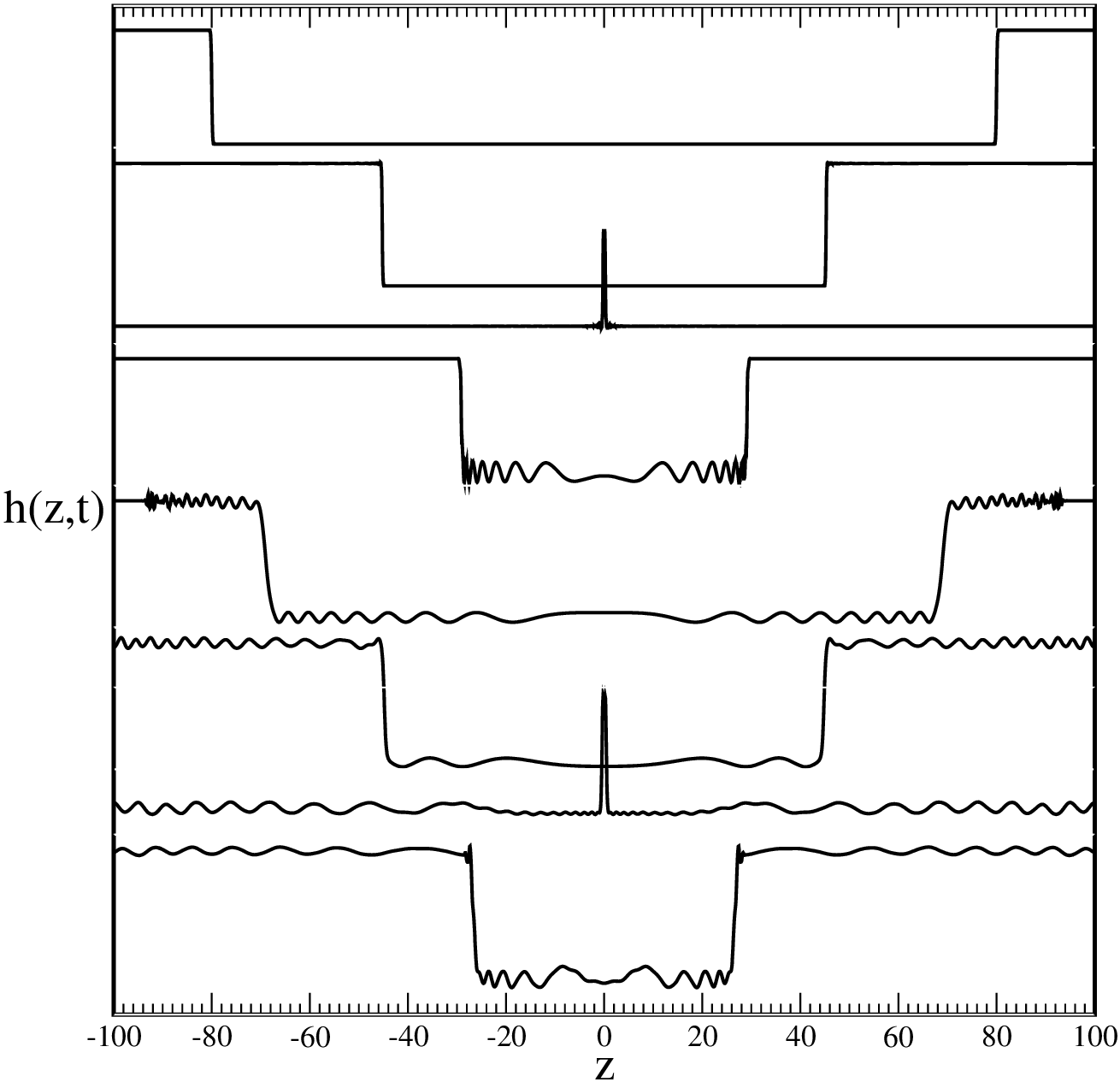} 
\includegraphics[width=0.49\textwidth, clip ]{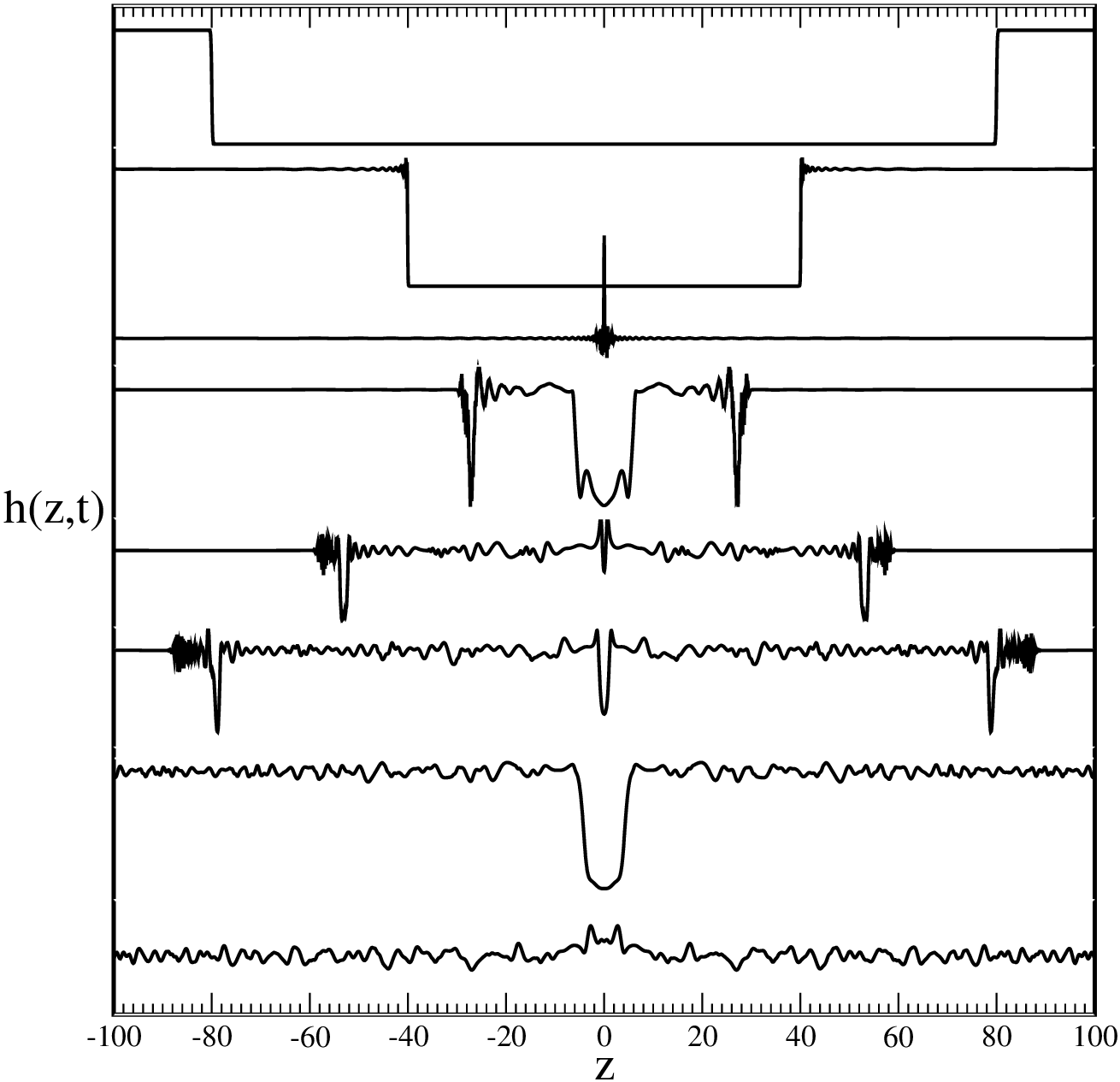} 
\caption{\small Snapshots of the field profile $h(z,t)$ during a bubble collision ($t$ increasing downwards). 
LEFT: Bubble collision for the potential with nearly degenerate minima (Figure 1 - Left). 
RIGHT: Bubble collision for the potential with very non-degenerate minima (Figure 1 - Right). In both cases, 
$\gamma_w = 10^2$, $l_w = 15/T_{\mathrm{EW}}$ and $T_{\mathrm{EW}} = 100$ GeV.}
\label{Fig:4}
\end{center}
\end{figure}

Following \cite{KS}, we compute the numerical solution for the field profile
$h(z,t)$ corresponding to the collision of two bubble walls, obtained from solving (\ref{eqmotion}) with a toy potential $V(h)$ 
of the form 

\be
V(h) = a^2 h^2 - b^2 h^3 + \lambda h^4
\ee 
 
\noindent both in the case of nearly degenerate minima (Figure 1 - Left) and very non-degenerate minima 
(Figure 1 - Right).
The results are shown in Figure \ref{Fig:4} (similar plots appeared earlier in \cite{joydivision}). 
Figure 2 - Left (corresponding to the 
potential of Figure 1 - Left) shows an approximately elastic bubble collision, 
while Figure 2 - Right (corresponding to the 
potential of Figure 1 - Right) shows a very inelastic one.

Guided by the numerical solution for $h(z,t)$ in the case of a very inelastic collision, we can 
obtain an analytic solution 
$h(z,t) = h_{\mathrm{TI}}$ for the case of a ``totally inelastic 
collision" (as opposed to the ``perfectly elastic collision" described earlier), in which
all the energy is radiated in the form of scalar waves after the bubble collision. For $t < 0$ 
(before the collision) we have 

\be
\label{inelastic1}
h_{\mathrm{TI}}(z,t < 0) = v_{T} + \frac{v_{T}}{2}
\left[\mathrm{Tanh}\left(\gamma_w\frac{z + t}{l_w}\right) - 
\mathrm{Tanh}\left(\gamma_w\frac{z - t}{l_w}\right) \right] 
\ee 

\noindent which matches $h_{l_{w}}(z,t < 0)$. In order to obtain $h_{\mathrm{TI}}(z,t)$ for $t > 0 $, we note 
that the field will 
not leave the basin of attraction of the broken minimum $v_T$ after the collision. We can then approximate 
the potential $V(h)$ the field will feel for $t > 0$ as 

\be
V(h) \simeq \frac{m_h^2}{2} \left(h - v_T  \right)^2 = \frac{m_h^2}{2} \delta h^2
\ee 

\noindent This allows to solve the equation of motion (\ref{eqmotion}) explicitly for 
$\delta h_{\mathrm{TI}}(z,t) \equiv h_{\mathrm{TI}}(z,t) - v_T$: 

\begin{eqnarray}
\label{eqmotioninelastic}
\left(\partial^2_t  - \partial^2_z \right) \delta h_{\mathrm{TI}}(z,t) = -m_h^2 \, \delta h_{\mathrm{TI}}(z,t) \nonumber \\
\, \nonumber \\
\delta h_{\mathrm{TI}}(z,0) = h_{l_{w}}(z,0) - v_T = 0 \\
\, \nonumber \\
\partial_t \, \delta h_{\mathrm{TI}}(z,0) = 
\frac{v_T\, \gamma_w}{l_w \left[ \mathrm{Cosh}\left(\frac{\gamma_w\, z}{l_w} \right) \right]^2} \nonumber 
\end{eqnarray}

\noindent where the boundary conditions follow from imposing continuity of $\delta h_{\mathrm{TI}}(z,t)$ and 
$\partial_t \, \delta h_{\mathrm{TI}}(z,t)$ at $t = 0$. From (\ref{eqmotioninelastic}), we finally obtain

\be
h_{\mathrm{TI}}(z,t > 0) = v_{T} \left[1 + \frac{l_w}{\gamma_w}
\int_{0}^{\infty} d p_z \,  \frac{p_z}{\sqrt{p_z^2 + m_h^2}} \,
\frac{\mathrm{Cos}\left(p_z\, z \right)}{\mathrm{Sinh} \left(\frac{\pi\, l_w \, p_z}{2\, \gamma_w} \right)} 
\, \mathrm{Sin}\left(\sqrt{p_z^2 + m_h^2}\, t \right) \right]
\ee 

Notice that in the limit $m_h \rightarrow 0$, (\ref{eqmotioninelastic}) becomes 
$\left(\partial^2_t  - \partial^2_z \right) \delta h_{\mathrm{TI}}(z,t) = 0$ and (\ref{inelastic1}) is also a solution for 
$t > 0$, case in which the two bubble walls would pass through each other without actually colliding. 

The analysis for the dynamics of bubble collisions presented here may be extended to phase transitions 
involving multiple fields (see for example \cite{Chinos}), although in this case the analysis of the field evolution 
after the bubble collision becomes much more complicated (since the scalar potential is multidimensional and the
field ``excursion" at the moment of the bubble collision will involve several fields), and we will not attempt 
it here.

\subsection{Particle Production Through Bubble Collisions}
\label{section22}

The bubble collision processes analyzed in the previous section allow to liberate into the plasma the energy 
contained in the bubble walls. This can happen either via direct particle production in the collisions
or via radiation of classical scalar waves which will subsequently decay into particles. For bubble collisions 
taking place in a thermal environment, the number densities $n_{\alpha}$ for the different particle species created 
during the collisions should very quickly approach the ones in thermal equilibrium $n^{\mathrm{EQ}}_{\alpha}$ after the phase transition,
thus rendering the particle production process irrelevant for the subsequent evolution of the Universe. 
However, as it has been briefly discussed in the introduction, under certain conditions 
fast thermalization of certain species after the phase transition 
may be avoided, which can make the particle production process very important in that case.

In order to study the particle production through bubble collisions, we will treat the 
scalar field configuration $h(z,t)$ as a classical external field and the states coupled to it as quantum 
fields in the presence of this source. In doing so, we will neglect the back-reaction of particle production 
on the evolution of the bubble walls themselves throughout the collision, which should be a good approximation
when the energy of the produced particles (for each species) is much less than the energy contained in the field configuration 
$h(z,t)$. The probability of particle production is given by \cite{Watkins:1991zt}

\be
\label{Number}
\mathcal{P} = 2 \, \mathrm{Im}\left(\Gamma\left[h\right]\right)   \quad \quad \quad \quad \left( \mathcal{P} \ll 1 \right)
\ee

\noindent where $\Gamma\left[ h \right]$ is the effective action. 
$\Gamma\left[ h \right]$ is the generating 
functional of 1PI Green functions, and to the quadratic order in $h$ 

\be
\label{EffAction}
\Gamma\left[ h \right] = \frac{1}{2} \int d^{4}x_1 \, d^{4}x_2 \, h(x_1) \,  h(x_2) \, \Gamma^{(2)} \left(x_1, x_2 \right)
\ee

\noindent with $\Gamma^{(2)} \left(x_1, x_2 \right) \equiv \Gamma^{(2)} \left(x_1 - x_2 \right)$ being the 2-point 1PI Green function. 
In terms of its Fourier transform $\tilde{\Gamma}^{(2)} \left(p^2\right)$, and using 
(\ref{Number}) and (\ref{EffAction}) we get

\be
\label{Number2}
\mathcal{P} = \int 
\frac{d^{4}p}{(2 \pi)^4} \mathrm{Im} \left( \tilde{\Gamma}^{(2)} \left(p^2 \right)
  \right) \int d^{4}x_1 \, d^{4}x_2 \, h(x_1) \,  h(x_2) \, e^{i p (x_1-x_2)}
\ee

The last integral in (\ref{Number2}) is just $\left| \tilde{h}(p) \right|^2$, with $\tilde{h}(p)$ being the Fourier transform
of the Higgs field configuration $h(x)$

\be
\label{Number3}
 \tilde{h}(p) = \int d^{4}x \, h(x) \, e^{i p \, x} 
\ee

For a background field configuration $h(z,t)$, its Fourier transform is given by $\tilde{h}(p) = 
(2 \pi)^2 \, \delta (p_x) \, \delta (p_y) \, \tilde{h}(p_z,\omega)$. Then, using (\ref{Number2}), we obtain
the mean number of particles produced per unit area \cite{Watkins:1991zt}:

\be
\label{Number4}
\frac{\mathcal{N}}{A} = 2 \int \frac{d p_z \, d \omega}{(2\,\pi)^2} \left| \tilde{h}(p_z,\omega) \right|^2
\mathrm{Im} \left( \tilde{\Gamma}^{(2)} \left(\omega^2 - p_z^2\right) \right)
\ee

The physical interpretation of (\ref{Number4}) is rather simple \cite{Watkins:1991zt}: the scalar field configuration 
$h(z,t)$, corresponding to the two bubble walls that approach and collide, can be decomposed into 
modes of definite four-momentum $p^2 = \omega^2 -p_z^2$ via the Fourier transform. Modes with $p^2>0$ represent 
propagating field quanta with mass squared $m^2 = p^2$. Then, (\ref{Number4}) integrates over the amount of field quanta 
of mass $p^2$ contained in the field configuration multiplied by the probability of those quanta to decay. 

The Fourier transform of the background field configuration $h(z,t)$ can be performed explicitly both for the case of a perfectly 
elastic collision and of a totally inelastic one analyzed in the previous section. For a perfectly elastic collision, in the limit 
of infinitely thin walls ($h(z,t) = h_{\infty}$), we obtain

\be
\label{2WallsFieldConfigurationFT}
\tilde{h}(p_z,\omega) = \tilde{h}_{\infty}(p_z,\omega) \equiv \frac{4\, v_T}{\omega^2- p_z^2}
\ee  

However, since the highest values of $p_z$ and $\omega$ available in the field configuration are naively expected to be of order 
$\gamma_w/l_w$ (modes with $p_z,\omega \gg  \gamma_w/l_w$ will be exponentially damped), the integration in (\ref{Number4}) 
should in this case be cut-off 
for $p_z > \gamma_w/l_w$ and $\omega > \gamma_w/l_w$. From (\ref{Number4}) and (\ref{2WallsFieldConfigurationFT}) 
we then obtain

\be
\label{Number5}
\frac{\mathcal{N}_{\infty}}{A} = \frac{32\, v_T^2}{\pi^2} \int_0^{\frac{\gamma_w}{l_w}} d \omega  
\int_0^{\frac{\gamma_w}{l_w}} d p_z \,
\frac{\mathrm{Im}\left(\tilde{\Gamma}^{(2)} \left(\omega^2 - p_z^2\right) \right)}{\left(\omega^2-p_z^2 \right)^2}
\ee

Alternatively, when the thickness of the bubble walls is accounted for ($h(z,t) = h_{l_w}$), the Fourier transform 
of (\ref{2WallsFieldConfiguration2}) gives

\be
\label{2WallsFieldConfigurationFT2}
\tilde{h}(p_z,\omega) = \tilde{h}_{l_w}(p_z,\omega) \equiv \frac{\pi \, l_w \, \omega}{2\,\gamma_w} \, 
\frac{4\, v_T}{\mathrm{Sinh} \left[ \frac{\pi \, l_w \, \omega}{2\, \gamma_w}\right]} \, \frac{1}{\omega^2-p_z^2}
\ee  

\noindent which automatically incorporates the exponential damping for $\omega, p_z \gg \gamma_w/l_w$. 
The mean number of particles per unit area now reads 

\be
\label{Number5Bis}
\frac{\mathcal{N}_{l_w}}{A} = \frac{8\, v_T^2  \, l_w^2}{\gamma_w^2} \int_0^{\infty} d \omega
\int_0^{\infty} d p_z \,
\frac{\mathrm{Im}\left(\tilde{\Gamma}^{(2)} \left(\omega^2 - p_z^2\right) \right)}{\left(\omega^2-p_z^2 \right)^2}
\frac{\omega^2}{\left( \mathrm{Sinh} \left[ \frac{\pi \, l_w \, \omega}{2\, \gamma_w}\right]  \right)^2}
\ee

For the opposite case of a totally inelastic collision ($h(z,t) = h_{\mathrm{TI}}$), the Fourier transform is given by 

\be
\label{2WallsFieldConfigurationFT3}
\tilde{h}(p_z,\omega) = \tilde{h}_{\mathrm{TI}}(p_z,\omega) \equiv \frac{\pi \, l_w \, p_z}{2\,\gamma_w} \, 
\frac{2\, v_T}{\mathrm{Sinh} \left[ \frac{\pi \, l_w \, p_z}{2\, \gamma_w}\right]} \,
\left(\frac{1}{\omega^2-p_z^2} - \frac{1}{\omega^2-p_z^2 - m_h^2} \right)
\ee  

The relative ``$- $" sign between the two contributions in (\ref{2WallsFieldConfigurationFT3}) 
can be easily understood noticing that in the limit $m_h \rightarrow 0$ the Fourier transform of $h_{\mathrm{TI}}(z,t)$ should give 
$\tilde{h}(p_z,\omega) \sim \delta (\omega \pm p_z)$. From (\ref{2WallsFieldConfigurationFT3}), the mean number 
of particles produced per unit area in the case of a totally inelastic collision is given by 

\be
\label{Number5BisBis}
\frac{\mathcal{N}_{\mathrm{TI}}}{A} = \frac{2\, v_T^2  \, l_w^2}{\gamma_w^2} \int_0^{\infty} d \omega 
\int_0^{\infty} d p_z \,
\frac{m_h^4 \, \mathrm{Im}\left(\tilde{\Gamma}^{(2)} \left(\omega^2 - p_z^2\right) \right)}
{\left(\omega^2-p_z^2 \right)^2 \left(\omega^2-p_z^2 - m_h^2\right)^2} 
\,  
\frac{p_z^2 }{\left( \mathrm{Sinh} \left[ \frac{\pi \, l_w \, p_z}{2\, \gamma_w}\right]  \right)^2}
\ee

The expressions (\ref{Number5}), (\ref{Number5Bis}) and (\ref{Number5BisBis}) can be rewritten in a more compact 
form by making the change of variables $\chi = \omega^2 - p_z^2$, $\Psi = \omega^2 + p_z^2$. 
After performing the integral in $\Psi$, the mean number of particles produced per unit area finally reads 

\be
\label{Number6}
\frac{\mathcal{N}}{A} = \frac{1}{2\, \pi^2} \int_0^{\infty} d \chi
\, f(\chi) \,
\mathrm{Im}\left(\tilde{\Gamma}^{(2)} \left( \chi \right) \right)
\ee
 
The function $f(\chi)$ encodes the details of the bubble collision process and quantifies the efficiency of particle production. 
For a perfectly elastic collision, in the limit of infinitely thin bubble walls, we have 

\be
\label{Number7}
f(\chi) = f_{\infty}(\chi) \equiv \frac{16\, v_T^2 \,
\mathrm{Log}\left[\frac{2\,\left(\frac{\gamma_w}{l_w}\right)^2 -\chi + 
2\,\frac{\gamma_w}{l_w} \sqrt{\left(\frac{\gamma_w}{l_w}\right)^2 - \chi} }{\chi}\right]}{\chi^2} \, \, 
\Theta \left[\left(\frac{\gamma_w}{l_w}\right)^2 - \chi \right]
\ee

For a perfectly elastic collision, and for bubble walls with finite thickness, we have 

\be
\label{Number7Bis}
f(\chi) = f_{l_w}(\chi) \equiv \frac{2\, \pi^2 \, l_w^2 \, v_T^2}{\gamma_w^2} \, \frac{1}{\chi^2} 
\int_{\chi}^{\infty} d\Psi \, \frac{\Psi +\chi}{\sqrt{\Psi^2 - \chi^2}} \, 
\frac{1}{\left( \mathrm{Sinh} \left[ \frac{\pi \, l_w \, \sqrt{\Psi + \chi}}{2\,\sqrt{2} \,\gamma_w}\right]  \right)^2}
\ee

Finally, for a totally inelastic collision, we have

\be
\label{Number7BisBis}
f(\chi) = f_{\mathrm{TI}}(\chi) \equiv \frac{\pi^2 \, l_w^2 \, v_T^2}{2 \, \gamma_w^2} \, 
\frac{m_h^4}{\chi^2 \left(\chi - m_h^2 \right)^2}
\int_{\chi}^{\infty} d\Psi \, \frac{\Psi -\chi}{\sqrt{\Psi^2 - \chi^2}} \, 
\frac{1}{\left( \mathrm{Sinh} \left[ \frac{\pi \, l_w \, \sqrt{\Psi - \chi}}{2\,\sqrt{2} \,\gamma_w}\right]  \right)^2}
\ee

\begin{figure}[ht]
\begin{center}
\includegraphics[width=0.49\textwidth, clip ]{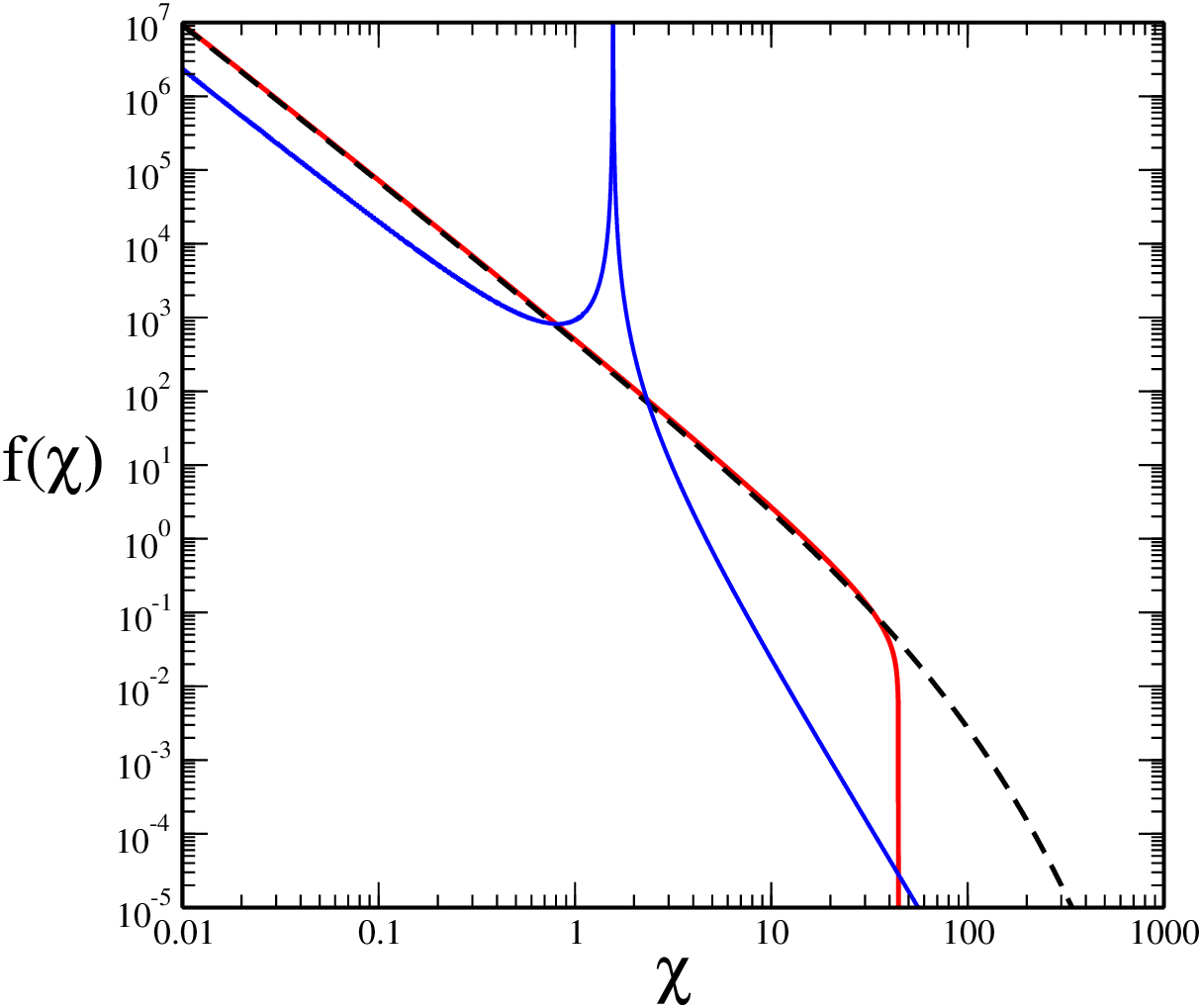}
\includegraphics[width=0.49\textwidth, clip ]{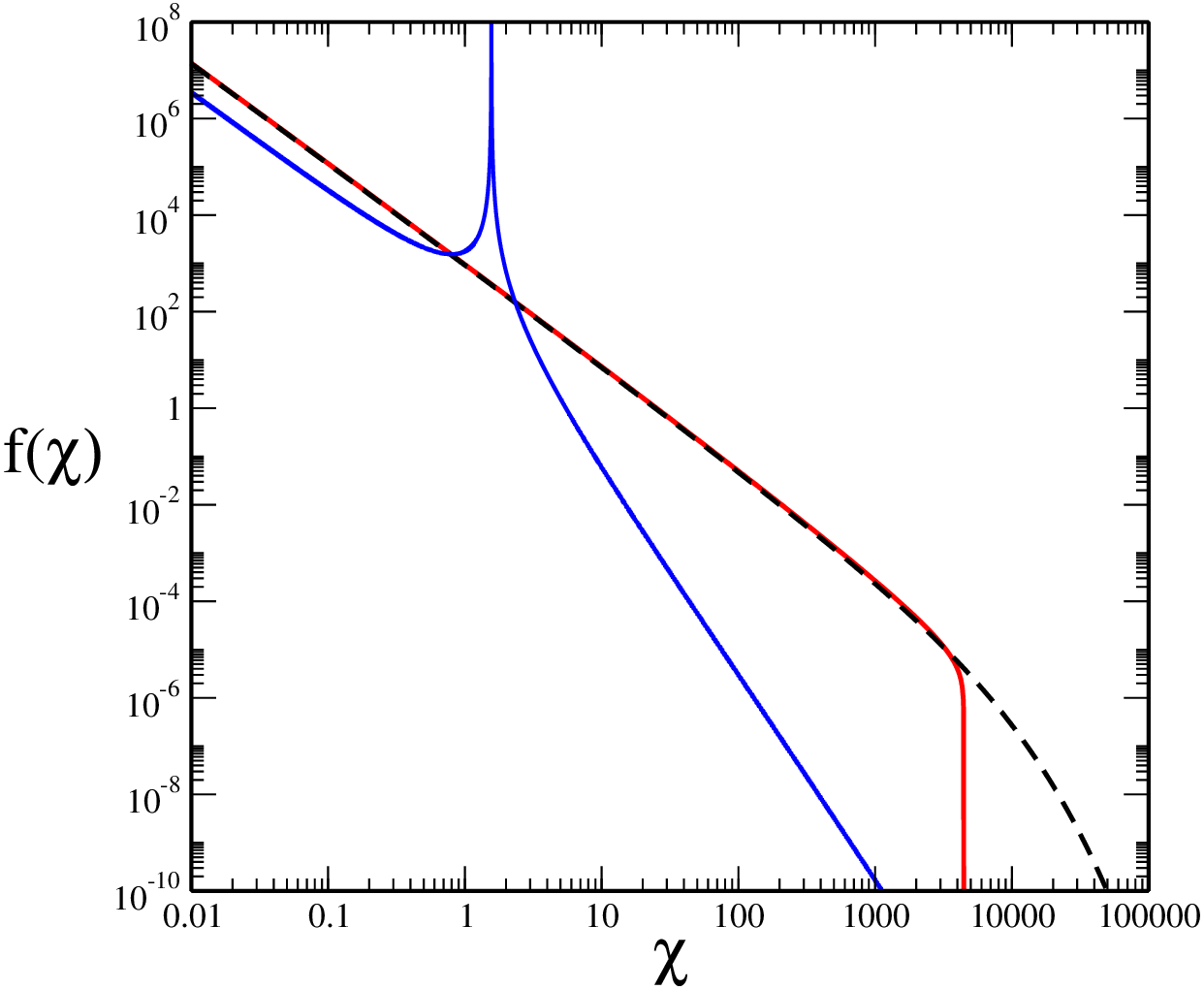}
\caption{\small Particle production efficiency $f(\chi \equiv \omega^2 - p_z^2)$ for $\gamma_w = 10^2$ (LEFT) and $\gamma_w = 10^3$
(RIGHT), $l_w = 15/T_{\mathrm{EW}}$ and $T_{\mathrm{EW}} = 100$ GeV, in the case of a perfectly
elastic collision with infinitely thin bubble walls (\ref{Number7}) (solid red) and with a finite bubble wall thickness 
(\ref{Number7Bis}) (dashed-black), and in the case of a totally inelastic collision (\ref{Number7BisBis}) (solid blue) with 
$m_h = 125$ GeV. The $\chi$-axis is displayed in units of $(100\,\, \mathrm{GeV})^2$.}
\label{Fig:5}
\end{center}
\end{figure}

In Figure \ref{Fig:5} we compare the efficiency $f(\chi)$ for the various cases (\ref{Number7}), (\ref{Number7Bis}) and 
(\ref{Number7BisBis}). Notice that $f_{\mathrm{TI}}(\chi)$ diverges as $\chi \rightarrow m_h^2$.
This divergence is artificial, due to considering $h(z,t)$ over infinite time and space, and should be cut-off since our solution 
is not valid over distances larger than the bubble radius $R_B$. 
Implementing this cut-off can be well approximated by replacing in (\ref{Number7Bis}) 

\be
\label{Number7Shift}
\left(\chi - m_h^2 \right)^2 \rightarrow \left(\chi - m_h^2 \right)^2 + (m_h^6 \,l_w^2)/\gamma_w^2.  
\ee

Defining $\chi_{\mathrm{min}}$ as the minimum value of $\chi$ for which particle production is possible (corresponding 
to the squared sum of the masses $M_{\alpha}$ of the particles being produced), we immediately see from 
Figure \ref{Fig:5} that for a totally inelastic collision, production of light particles ($\chi_{\mathrm{min}} < m_h^2$) 
may be very efficient, while production of heavy particles ($\chi_{\mathrm{min}} \gg m_h^2$) will be extremely suppressed. 
For a perfectly elastic collision, however, the production of heavy particles may be relatively efficient (we will
comment further on this point at the end of section \ref{section23}).  
For the study of the efficiency of particle production in varios different scenarios in the next sections, we will use 
(\ref{Number7}) for the case of an elastic collision, while for the case of a very inelastic one it is possible to show that 
(\ref{Number7BisBis}) (together with (\ref{Number7Shift})) can be approximated as   

\be
\label{Number7BisBisBis}
f_{\mathrm{TI}}(\chi) \simeq 4 \, v_T^2\, m_h^4 \, \frac{
\mathrm{Log}\left[\frac{2\,\left(\frac{\gamma_w}{l_w}\right)^2 +\chi + 
2\,\frac{\gamma_w}{l_w} \sqrt{\left(\frac{\gamma_w}{l_w}\right)^2 + \chi} }{\chi}\right]}{\chi^2 
\left[\left(\chi - m_h^2 \right)^2 + m_h^6 \frac{l_w^2}{\gamma_w^2} \right]} \, .
\ee

Let us now turn to the evaluation of the imaginary part of the 2-point 1PI Green function's Fourier transform 
$\tilde{\Gamma}^{(2)} \left(\chi \equiv \omega^2 - p_z^2\right)$. Through the optical theorem, we can write:

\be
\label{N/A2}
\mathrm{Im}\left(\tilde{\Gamma}^{(2)} \left(\chi \right) \right) = 
\frac{1}{2} \sum_{\alpha} \int d\Pi_{\alpha} \left|\overline{\mathcal{M}}(h \rightarrow \alpha) \right|^2 
\, \Theta \left[\chi- \chi_{\mathrm{min}} \right]
\ee

\noindent where $\left|\overline{\mathcal{M}}(h\rightarrow \alpha) \right|^2$ is the spin-averaged squared amplitude 
for the decay of $h$ into a set of particles $\alpha$ with masses $M_{\alpha}$, $\chi_{\mathrm{min}} \equiv \left(\sum M_{\alpha}\right)^2$ 
is the minimum value of $\chi$ for which this decay is possible and $d\Pi_{\alpha}$
is the \textit{relativistically invariant} $n$\textit{-body phase space} element

\be
d\Pi_{\alpha} = \left(\prod_i \frac{d^3\,k_i}{(2\,\pi)^3} \frac{1}{2\,E_i}\right) (2\,\pi)^4 \, \delta^4(p-\sum_i k_i) 
\ee

Then, the number of particles of a certain type $\alpha$ produced per unit area during the bubble collision will simply read
from (\ref{Number6}) and (\ref{N/A2})

\be
\label{N/A3}
\left.\frac{\mathcal{N}}{A}\right|_{\alpha} = 
\frac{1}{4\, \pi^2} \int_{\chi_{\mathrm{min}}}^{\infty} d \chi
\, f(\chi) \, \int d\Pi_{\alpha} \left|\overline{\mathcal{M}}(h\rightarrow \alpha) \right|^2
\ee

The amount of energy produced per unit area in the form of particles $\alpha$ is obtained by weighting (\ref{N/A3})
by the energy of each decaying Fourier mode. This yields

\be
\label{E/A1}
\left.\frac{\mathcal{E}}{A}\right|_{\alpha} = 
\frac{1}{4\, \pi^2} \int_{\chi_{\mathrm{min}}}^{\infty} d \chi
\, f(\chi) \sqrt{\chi} \, \int d\Pi_{\alpha} \left|\overline{\mathcal{M}}(h\rightarrow \alpha) \right|^2
\ee
  
From (\ref{N/A3}) and (\ref{E/A1}), the non-thermally produced energy density $\rho_{\alpha}$ 
(assuming that the produced particles quickly diffuse into the bubble interior)  reads 

\be
\label{Energydensity}
\rho_{\alpha} \equiv \left.\frac{\mathcal{E}}{V}\right|_{\alpha} = \left.\frac{\mathcal{E}}{A}\right|_{\alpha} \, \frac{A}{V} 
\simeq \left.\frac{\mathcal{E}}{A}\right|_{\alpha} \, \frac{3}{2 \, R_B}
\ee

\noindent with $A \sim 4 \,\pi\, R_B^2$ being the total collision area and $V$ the volume of the two colliding bubbles. 
From (\ref{Energydensity}), and bearing in mind that $R_B \simeq \beta^{-1}$, the non-thermally generated comoving energy density is 

\be
\label{ComEnergydensity}
\Upsilon_\alpha  = \frac{\rho_{\alpha}}{s(T_{\mathrm{EW}})} \simeq  
\frac{20}{\sqrt{\pi\, g_*}} \, \frac{1}{M_{\mathrm{Pl}}\, T_{\mathrm{EW}}} \, \frac{\beta}{H} \, 
\left.\frac{\mathcal{E}}{A}\right|_{\alpha} 
\ee

\noindent with $s(T_{\mathrm{EW}})$ the entropy density after the EW phase transition.

\section{Particle Production via the Higgs Portal}
\label{section23}

The efficiency of particle production may strongly depend on the nature of the particles being produced. 
In this section we will analyze the particle production efficiency for scalars $S$, fermions $f$ and vector bosons 
$V_{\mu}$ coupled to the Higgs field. Apart from estimating the production of SM fermions and gauge bosons through
this process, we will consider a simple Higgs-portal extension of the SM in order to study the production of other
possible scalar, fermion or vector boson particles. Furthermore, we will restrict ourselves to $Z_2$ symmetric 
Higgs-portal scenarios, since we will ultimately be interested in dark matter analyses. 
We also comment on how to interpret the results in the case when the calculated particle production 
exceeds the energy available in the bubble wall.  

\subsection{Scalars}
\label{section231}

For the complex scalar $S$ interacting with the SM via the Higgs portal, the relevant part of the 
lagrangian is given by 

\be
\label{scalarsLagrangian}
-\Delta \mathcal{L}_s = m_s^2\, |S|^2
+ \lambda_s \left|H\right|^2 \, |S|^2    \quad \quad \quad \mathrm{with} \quad H = 
\left(\begin{array}{c}
 0 \\
\frac{h + v_T}{\sqrt{2}}
\end{array}\right).
\ee 
 
\begin{figure}[ht]
\begin{center}
\includegraphics[width=0.49\textwidth, clip ]{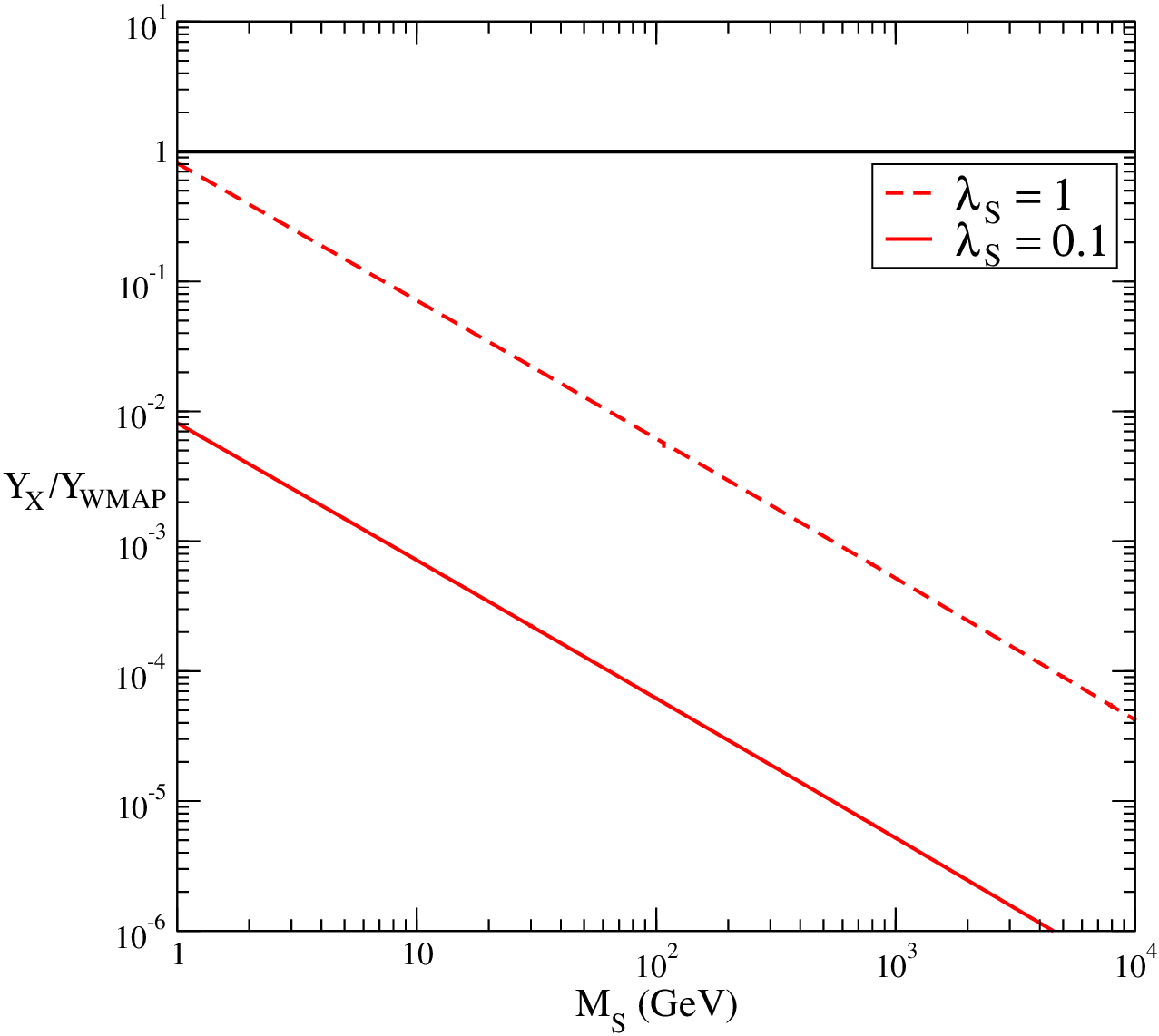}
\includegraphics[width=0.49\textwidth, clip ]{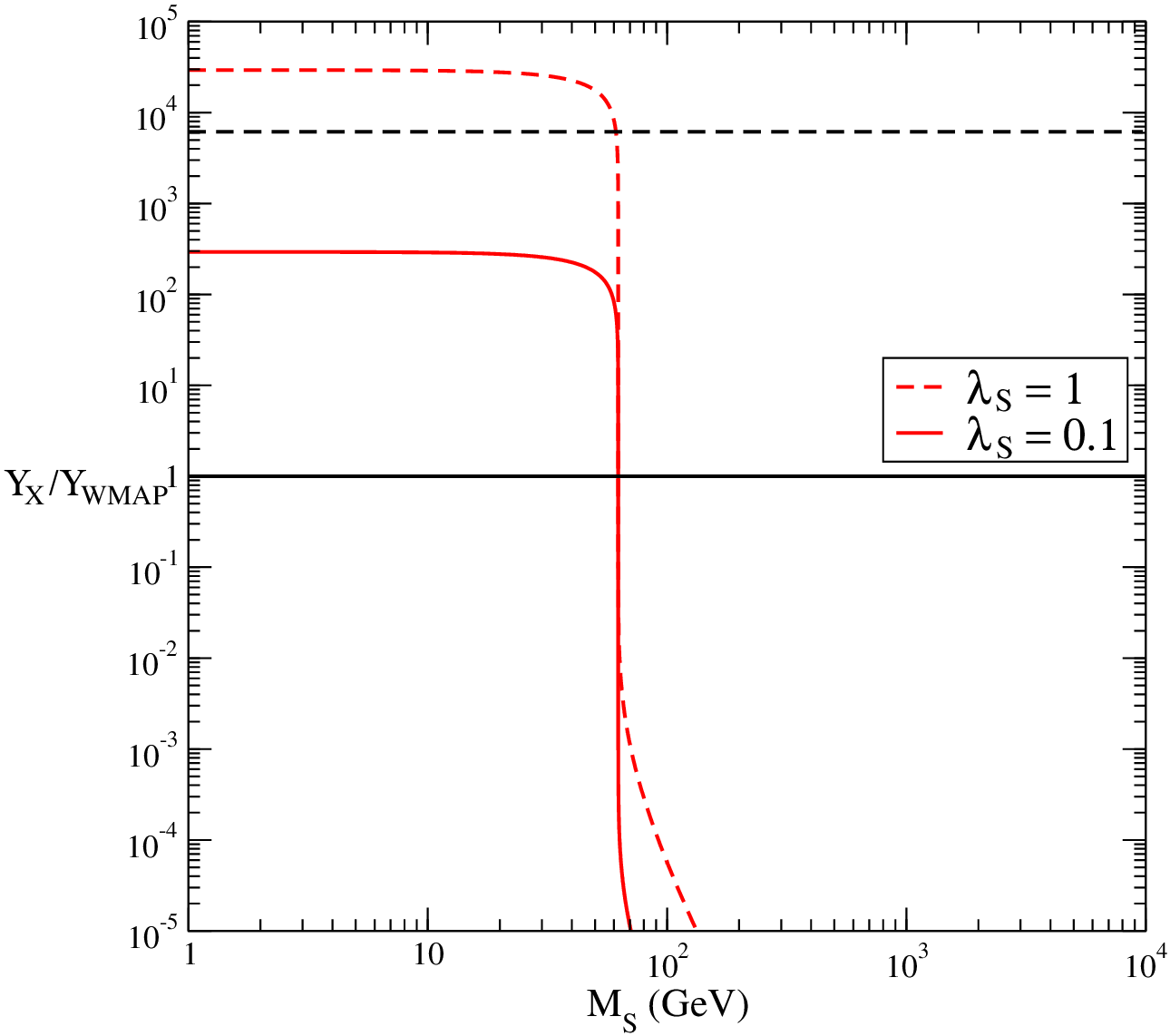}
\caption{\small Generated comoving energy density $\Upsilon$ in scalars $S$ (normalized to the observed dark 
matter comoving energy density) as a function of the scalar mass $M_s$ in the 
perfectly elastic collision limit (LEFT) and totally inelastic collision limit (RIGHT) for $\gamma_w = 10^8$,
$l_w = 15/T_{\mathrm{EW}}$ and $T_{\mathrm{EW}} = 100$ GeV. The solid-black line corresponds to the 
observed dark matter comoving energy density, and the dashed-black line (not seen in LEFT) corresponds 
to the maximum possibly generated comoving energy density ($\mathcal{E} = E_w$).}
\label{Fig:9}
\end{center}
\end{figure}

In this case, $\left|\mathcal{M}(h \rightarrow S \, \bar S) \right|^2 = \lambda_s^2 \, v_T^2$, and one immediately obtains 

\be
\label{scalarsProd1}
\mathrm{Im}\left[\tilde{\Gamma}^{(2)} \left( \chi \right) \right]_{S} = 
\lambda_s^2 \, v_T^2 \int d\Pi_{S} = \sqrt{1-4 \frac{M_{s}^2}{\chi}}\, \frac{\lambda_s^2 \, v_T^2}{8 \pi}\, \,
\Theta \left(\chi- 4 M_{s}^2\right) 
\ee

\noindent with $M_s^2 \equiv m_s^2 + (\lambda_s/2)\, v_T^2$ being the scalar squared mass. Then, 
using (\ref{Number6}), (\ref{Number7}), (\ref{Number7BisBisBis}), (\ref{ComEnergydensity}) and 
(\ref{scalarsProd1}) we can compute the $S$-scalar comoving energy density generated through 
the bubble collisions (normalized to the observed dark matter comoving energy 
density) as a function of $M_s$ and $\lambda_s$. The results are shown in Figure \ref{Fig:9}.

From Figure \ref{Fig:9} it can be clearly seen that scalar particle production is quite suppressed 
for elastic collisions. For very inelastic collisions, heavy-scalar particle production is extremely suppressed, 
while production of light scalars turns out to be very efficient in this case. In fact, Figure \ref{Fig:9} shows 
that for large values of $\lambda_s$ ($\lambda_s \lesssim 1$) the naively calculated energy of the produced particles 
$\mathcal{E}$ exceeds the amount of energy on the bubble walls $E_w$.  
 That inconsistency indicates  that in these cases backreaction cannot be neglected. 
We will comment  and expand on this issue in section \ref{section24}.   

\subsection{Fermions}
\label{section232}

Turning now to fermionic particle production, in the presence of a tree-level 
Yukawa coupling between the Higgs and the fermions
$\lambda_{f} H\, \overline{f} \, f$, the squared decay amplitude reads

\be
\label{fermionsProd1}
\left|\overline{\mathcal{M}}(h \rightarrow \overline{f}\, f) \right|^2
= 2\, \lambda_{f}^2 \left(p^2 - 4\, m_{f}^2 \right)
\ee

\noindent which, in the case of SM fermions, leads directly to 

\be
\label{fermionsProd2}
\mathrm{Im}\left[\tilde{\Gamma}^{(2)} \left( \chi \right) \right]_{f} = 
\frac{m_f^2}{4 \pi\,v_T^2} \, \,\chi \,\left(1-\frac{4\,m_{f}^2}{\chi} \right)^{\frac{3}{2}} \,\,
\Theta \left(\chi- 4 m_{f}^2\right) 
\ee

\begin{figure}[ht]
\begin{center}
\includegraphics[width=0.49\textwidth, clip ]{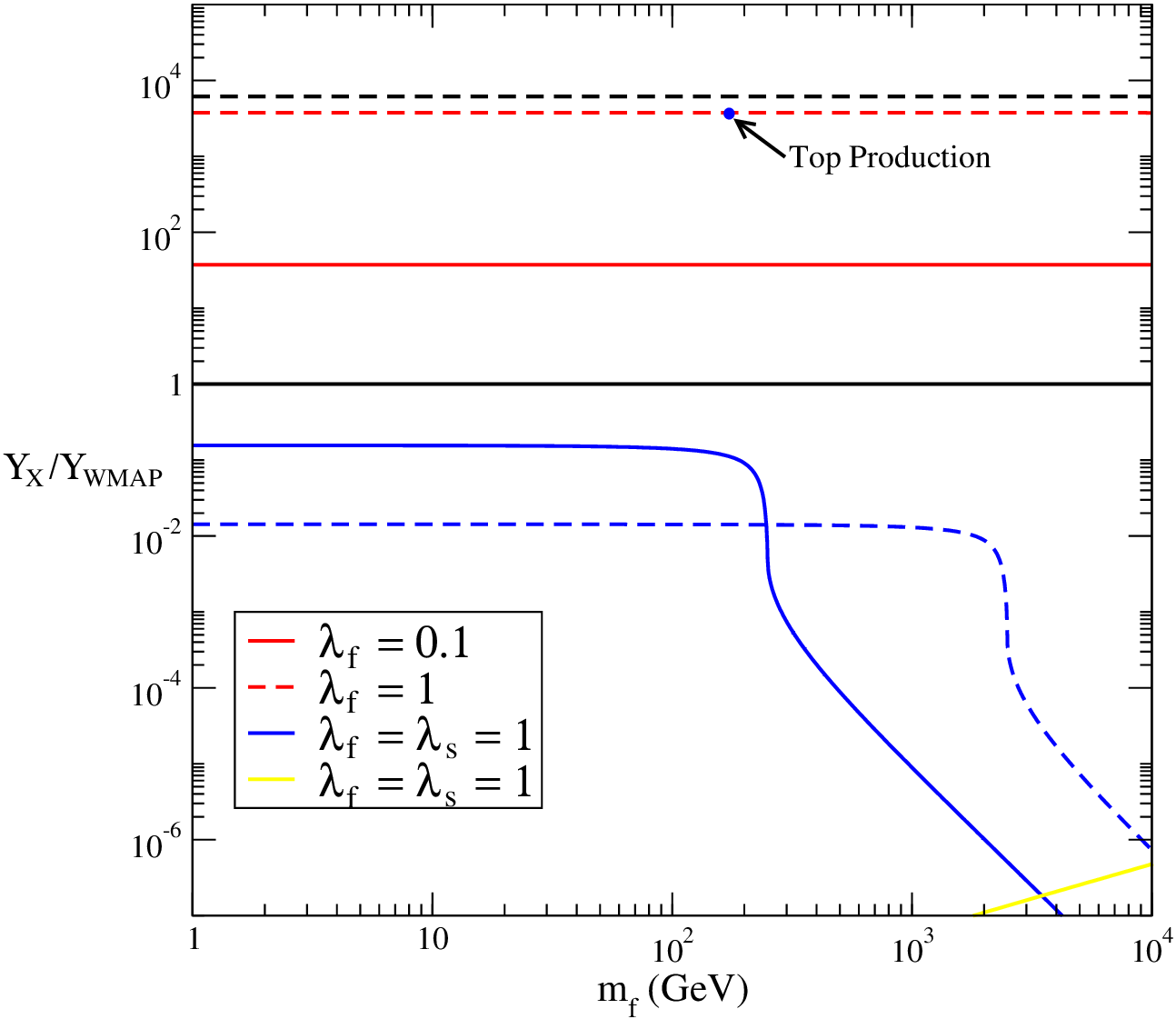}
\includegraphics[width=0.49\textwidth, clip ]{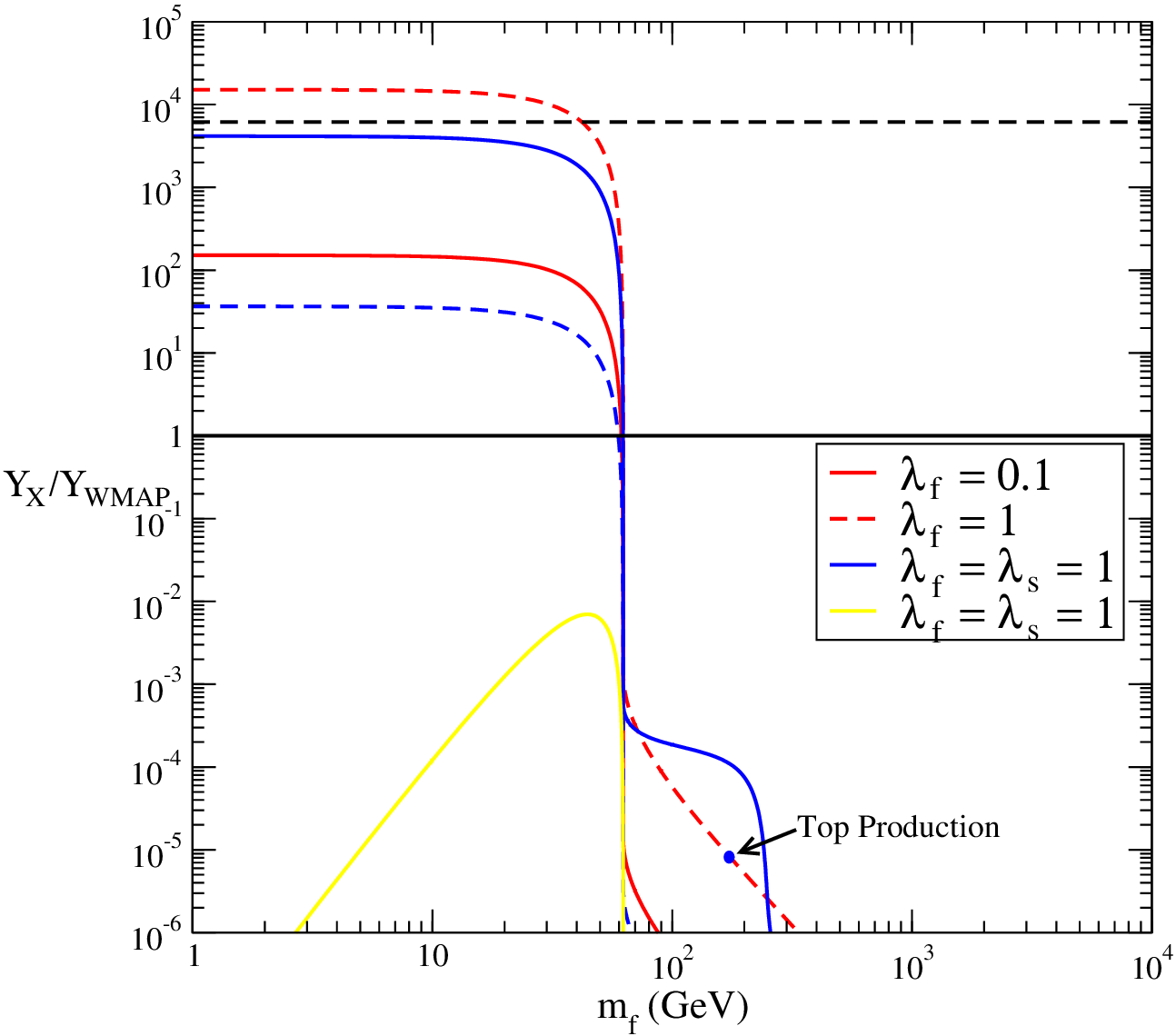}
\caption{\small Generated comoving energy density $\Upsilon$ in fermions $f$ (normalized to the observed dark 
matter comoving energy density) as a function of the fermion mass $m_f$ in the 
perfectly elastic collision limit (LEFT) and totally inelastic collision limit (RIGHT) for $\gamma_w = 10^8$,
$l_w = 15/T_{\mathrm{EW}}$ and $T_{\mathrm{EW}} = 100$ GeV. Red lines: production 
in the presence of a direct tree-level Yukawa coupling between fermions and Higgs (\ref{fermionsProd2}). 
Blue lines: production for a tree-level effective coupling (\ref{fermionsProd5}), 
for $\mu_s = M_s=500$ GeV (solid) and $5$ TeV (dashed). Yellow lines: production for a 1-loop effective coupling
(\ref{fermionsProd6}). The solid-black line corresponds to the observed dark matter comoving
energy density, and the dashed-black line corresponds to the maximum possible generated comoving number density 
($\mathcal{E} = E_w$).}
\label{Fig:10}
\end{center}
\end{figure}

The production of (SM) fermions will then be enhanced with respect to the one of 
Higgs-portal $S-$scalars (specially in the limit of very elastic collisions, see 
Figure \ref{Fig:10}) due to the extra factor $\left(\chi - 4\, m_{f}^2 \right)$ in (\ref{fermionsProd2}). 
Scenarios where the fermionic particle production might be important include (apart from the SM itself) 
the MSSM and its various extensions, due to the tree-level coupling between 
Higgses, Higgsinos and Gauginos\footnote[4]{In particular, the production of neutralino dark matter 
might have an impact on the subsequent evolution of the Universe.}. 
 
In the absence of a direct coupling, the interaction between the Higgs and the fermions will 
occur via an effective operator. This is the case for the so-called fermionic Higgs-portal:

\be
\label{fermionsLagrangian}
-\Delta \mathcal{L}_{f} = m_{f}\, \overline{f} f
+ \frac{\lambda_{f}}{\Lambda} \left|H\right|^2 \, \overline{f} f
\ee

However, since bubble collisions may excite very massive Higgs field modes 
($p^2 \gg T_{\mathrm{EW}}^2$), particle production in this case may be sensitive to 
the $\mathrm{UV}$ completion of the Higgs-portal effective theory, making it unreliable 
to compute the particle production in the 
fermionic Higgs-portal via (\ref{fermionsLagrangian}). 
Here we consider a simple $\mathrm{UV}$ completion for the fermionic Higgs-portal,
and compute the particle production in this case. We add a singlet scalar field $S$
as a mediator between the Higgs field and the fermion $f$, the relevant part of the lagrangian being

\be
\label{fermionsLagrangian2}
-\Delta \mathcal{L}_{f} = \frac{m_s^2}{2}\, S^2
+ \frac{\lambda_s}{2} \left|H\right|^2 \, S^2 +\mu_s \left|H\right|^2 \, S + m_{f}\, \overline{f} f
+ \lambda_{f} S \, \overline{f} f
\ee

For simplicity, we will avoid a vev for $S$ (it can be done through the addition of a linear term for $S$ in 
(\ref{fermionsLagrangian2})). For $\mu_s \neq 0$ the effective fermionic Higgs-portal operator 
$\left|H\right|^2 \,\overline{f} f $ will be generated at tree-level. The squared decay amplitude for 
$h \rightarrow \overline{f}\, f$ will then be

\be
\label{fermionsProd3}
\left|\overline{\mathcal{M}}(h \rightarrow \overline{f}\, f) \right|^2
= 2\, \frac{\lambda_{f}^2\,\mu_s^2\,v_T^2}{\left(p^2 - M_{s}^2\right)^2 + 
\Gamma^2_s\,M_{s}^2} \left(p^2 - 4\, m_{f}^2 \right)
\ee

\noindent with

\be
\label{fermionsProd4}
\Gamma_s = \frac{\lambda_{s}^2\,v_{T}^2 + \mu_s^2}{16\,\pi\,M_{s}} \sqrt{1- \frac{4\,m_{h}^2}{M_{s}^2}} \,\, 
\Theta \left( M^2_{s} - 4\, m_{h}^2 \right) +
\frac{\lambda_{f}^2\, M_{s}}{8\,\pi} \left(1- \frac{4\,m_{f}^2}{M_{s}^2}\right)^{\frac{3}{2}} \,\, 
\Theta \left( M^2_{s} - 4\, m_{f}^2 \right)
\ee

\noindent leading finally to

\be
\label{fermionsProd5}
\mathrm{Im}\left[\tilde{\Gamma}^{(2)} \left( \chi \right) \right]_{f} = 
\frac{\lambda_{f}^2\,\mu_s^2\,v_T^2}{4 \pi} \, \frac{\chi}{\left(\chi - M_{s}^2\right)^2 + \Gamma^2_s\,M_{s}^2}
\, \left(1-\frac{4\,m_{f}^2}{\chi}\right)^{\frac{3}{2}} 
\Theta \left(\chi- 4 m_{f}^2\right) 
\ee

When $\mu_s = 0$ the effective fermionic Higgs-portal operator
is not generated at tree-level, but rather the decay 
$h \rightarrow \overline{f} \, f$ occurs via a finite 1-loop diagram, yielding

\be
\label{fermionsProd6}
\mathrm{Im}\left[\tilde{\Gamma}^{(2)} \left( \chi \right) \right]_{f} = 
\frac{\left(\lambda_s \, \lambda_{f}^2\right)^2}{\left(4 \pi\right)^5} 
\, F \left[m_{f}^2,\, M_{s}^2,\, \chi \right]\, \chi \, \left(1-\frac{4\,m_{f}^2}{\chi} \right)^{\frac{3}{2}}\, \,
\Theta \left(\chi- 4 m_{f}^2\right) 
\ee

\noindent where $F \left[m_{f}^2,\, M_{s}^2,\, \chi \right]$ is a form factor that scales as 

\be
\label{fermionsProd7}
F \left[m_{f}^2,\, M_{s}^2,\, \chi \right] \, \longrightarrow \, 
\frac{m_{f}^4}{\chi^{2}} \,  \mathrm{Log}\left(\frac{\chi}{m_{f}^2}\right) \quad \quad \quad \chi 
\gg m_{f}^2,\, M_{s}^2
\ee

Fermionic Higgs-portal particle production both in the $\mu_s = 0$ and $\mu_s \neq 0$ is shown in 
Figure \ref{Fig:10}, where it can be clearly seen that the production in the absence of a direct 
coupling between the Higgs and the fermions $f$ differs from what would have been naively obtained 
using (\ref{fermionsLagrangian}). As for the case of scalar particle production, under certain 
circumstances the estimate of fermionic particle production neglecting backreaction exceeds the 
amount of energy stored in the bubble walls ($\mathcal{E} > E_w$), and in order to obtain a 
physically meaningful result backreaction should be included (We will expand on this issue in section 
\ref{section24}).

\subsection{Vector Bosons}
\label{section233}

Finally, we study the production of vector boson particles. In the presence of a tree-level 
coupling between the Higgs and the vector bosons $\lambda_{V} M_{V} \, h \, V_{\mu} V_{\mu}$, 
the squared decay amplitude reads

\be
\label{VectorProd1}
\left|\overline{\mathcal{M}}(h \rightarrow V_{\mu}\,V_{\mu}) \right|^2
= \lambda^2_{V} M^2_{V} \left(3 - \frac{p^2}{M^2_{V}} + \frac{p^4}{4\,M^4_{V}} \right)
\ee

\noindent leading to 

\be
\label{VectorProd2}
\mathrm{Im}\left[\tilde{\Gamma}^{(2)} \left( \chi \right) \right]_{V} = 
\frac{\lambda^2_{V} M^2_{V}}{8 \pi} \, \left(3 - \frac{\chi}{M^2_{V}} + \frac{\chi^2}{4\,M^4_{V}} \right) 
\sqrt{1-4 \frac{M_{V}^2}{\chi}}\, \,
\Theta \left(\chi- 4 M_{V}^2\right) 
\ee

\begin{figure}[ht]
\begin{center}
\includegraphics[width=0.49\textwidth, clip ]{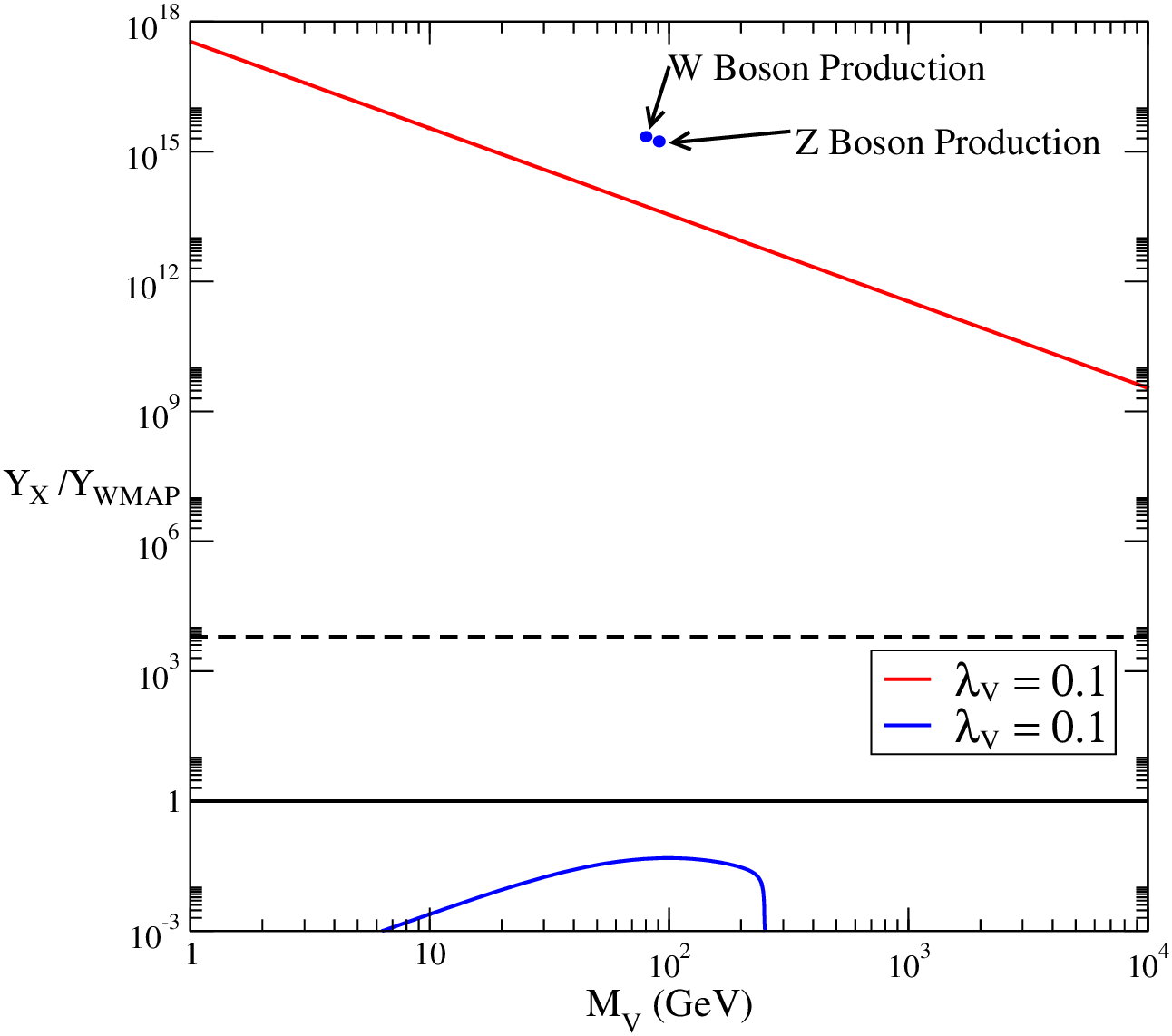}
\includegraphics[width=0.49\textwidth, clip ]{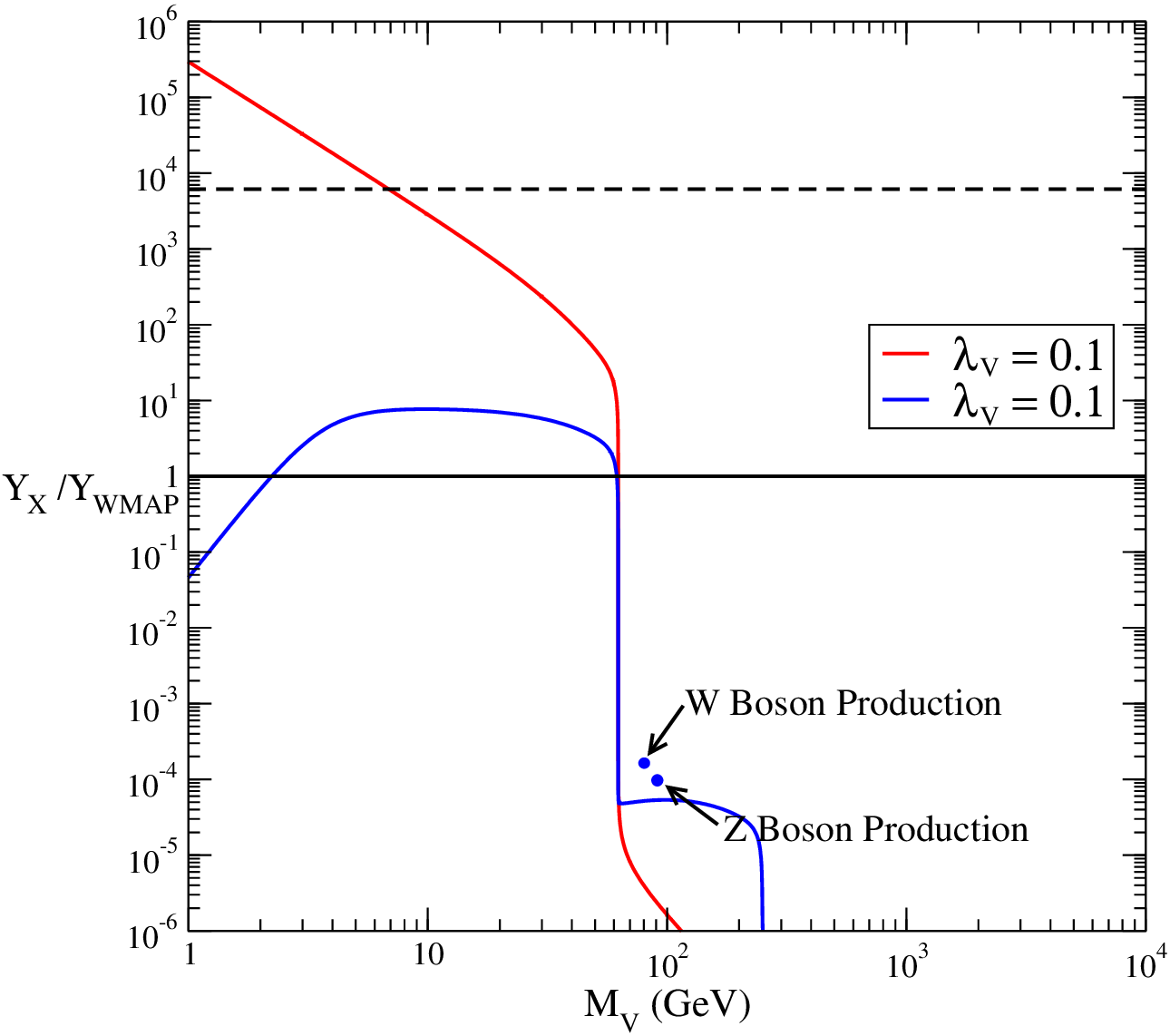}
\caption{\small 
Generated comoving energy density $\Upsilon$ in vector bosons $V_{\mu}$ (normalized to the observed dark 
matter comoving energy density) as a function of the vector boson mass $M_{V}$ in the 
perfectly elastic collision limit (LEFT) and totally inelastic collision limit (RIGHT) for $\gamma_w = 10^8$,
$l_w = 15/T_{\mathrm{EW}}$ and $T_{\mathrm{EW}} = 100$ GeV. Red line: production 
in the presence of a direct tree-level coupling between vector bosons and Higgs (\ref{VectorProd2}). 
Blue line: production for a tree-level effective coupling (\ref{VectorProd5}), 
for $\lambda_{hs} = 1$ and $M_s = 500$ GeV. The solid-black line corresponds to the observed dark matter comoving
energy density, and the dashed-black line corresponds to the maximum possible generated comoving number density 
($\mathcal{E} = E_w$).}
\label{Fig:11}
\end{center}
\end{figure}

Comparing (\ref{scalarsProd1}), (\ref{fermionsProd2}) and (\ref{VectorProd2}) we immediately 
observe the relative efficiency of particle production for scalars, fermions and vector bosons. While 
$\mathrm{Im}\,[\tilde{\Gamma}^{(2)} \left( \chi \right)]$ scales as $\chi^0$ for scalars, 
and as $\chi$ for fermions, in the case of vector bosons it scales as $\chi^2$, thus greatly enhancing 
production of vector bosons with respect to scalars or fermions for very elastic collisions 
(see Figure \ref{Fig:11}). It is then expected that most of the available energy from the EW
phase transition will go into $W_{\mu}$ and $Z_{\mu}$ gauge boson production and (possibly) 
other vector bosons coupled at tree-level to the Higgs in extensions of the 
SM\footnote[5]{such as Little Higgs theories or extra-dimensional 
scenarios with gauge fields living in the bulk.}.

In the absence of a direct coupling, the interaction between the Higgs and the vector bosons may 
occur via an effective operator, as in the so-called vector Higgs-portal \cite{Lebedev:2011iq}:

\be
\label{vectorsLagrangian}
-\Delta \mathcal{L}_{V} = \frac{1}{2} m_{V}^2\, V_{\mu}V^{\mu}
+ \lambda_{V} \left|H\right|^2 \, V_{\mu}V^{\mu}
\ee

However (like for the fermionic Higgs-portal) an analysis of vector boson particle production 
in the context of the effective theory (\ref{vectorsLagrangian}) will be unreliable due to 
very massive Higgs field modes ($p^2 \gg T_{\mathrm{EW}}^2$) being excited during the bubble 
collisions. Vector boson particle production will then be sensitive to the way in which the 
effective operator $\left|H\right|^2 \, V_{\mu}V^{\mu}$ is generated. One possible way of generating 
the effective operator at tree-level, being $V_{\mu}$ a hidden $U(1)$ gauge field, is by 
integrating out a $U(1)$-charged complex scalar $S$ which has a Higgs portal coupling 
$\left|H\right|^2 \, S^* S$, the relevant part of the lagrangian then being

\be
\label{vectorsLagrangian2}
-\Delta \mathcal{L}_{V} = \frac{1}{4}F_{\mu\nu}F^{\mu\nu} - D_{\mu}S^{*}D^{\mu}S + V(S) + \lambda_{hs}\,
\left|H\right|^2 \, S^*S
\ee

In this scenario, the vector boson $V_{\mu}$ acquires a mass via the spontaneous breaking of the hidden $U(1)$, 
through a vev $v_{S}$ for the $S$-scalar\footnote[6]{This implies that there may have been another phase 
transition in the early Universe associated with the spontaneous breaking of the hidden $U(1)$ 
gauge symmetry, which we need to require to have happened long before the EW phase transition 
since otherwise the EW phase transition would have been effectively multi-field and our present 
analysis of particle production would be totally unrealistic.}. The squared decay amplitude for 
$h \rightarrow V_{\mu}\, V_{\mu}$ will then be

\be
\label{VectorProd3}
\left|\overline{\mathcal{M}}(h \rightarrow V_{\mu}\, V_{\mu}) \right|^2
= \frac{\lambda_{hs}^2}{4} \, \frac{v_T^2\, M^4_{V}}{\left(p^2 - M_{s}^2\right)^2 + \Gamma^2_s\,M_{s}^2}
\,\left(3 - \frac{p^2}{M^2_{V}} + \frac{p^4}{4\,M^4_{V}} \right)
\ee

\noindent with $\Gamma_s$ being the decay width of $S$. This leads to 

\be
\label{VectorProd5}
\mathrm{Im}\left[\tilde{\Gamma}^{(2)} \left( \chi \right) \right]_{V} = 
\frac{\lambda_{hs}^2}{32 \,\pi} \, \,
\frac{v_T^2\,M^4_{V}
\left(3 - \frac{\chi}{M^2_{V}} + \frac{\chi^2}{4\,M^4_{V}} \right)}{\left(\chi - M_{s}^2\right)^2 + 
\Gamma^2_s\,M_{s}^2}
\,\sqrt{1- \frac{4\,M_{V}^2}{\chi}} 
\, \,
\Theta \left(\chi- 4 M_{V}^2\right) 
\ee

Vector boson effective Higgs-portal particle production is shown in Figure \ref{Fig:11}, resulting in a very 
suppressed particle production with respect to the case in which the vector bosons and the Higgs couple 
directly at tree-level, specially for very elastic collisions. From Figure \ref{Fig:11} it is also clear 
that backreaction is most important for direct vector boson particle production (for which the production 
estimate yields $\mathcal{E} \gg E_w$).

\subsection{Backreaction and Relative Efficiency}
\label{section24}

Clearly, for the present analysis of particle production to be physically meaningful it must be assumed that 
the total energy of the produced particles is less than the energy contained in the background field 
configuration $h(z,t)$. Moreover, when the energy of the produced particles starts being comparable to 
the energy of the background field we expect backreaction on $h(z,t)$ due to the particle production
to be important. Then, in order for the previous analysis to be reliable, it is needed

\be
\label{Backreaction}
\left.\frac{\mathcal{E}}{A}\right|_X \ll 
\frac{E_w}{A} = \frac{2}{3}\, v_{T}^2 \,\frac{\gamma_w}{l_w}
\ee 

As it has been shown in the previous section, for fermion or vector boson particle production 
the previous condition (\ref{Backreaction}) is not satisfied, and in some cases 
even $\mathcal{E} \gg E_w$ is obtained (Figure \ref{Fig:11} LEFT), 
signaling the extreme importance of backreaction in those scenarios. 

\begin{figure}[ht]
\begin{center}
\includegraphics[width=0.49\textwidth, clip ]{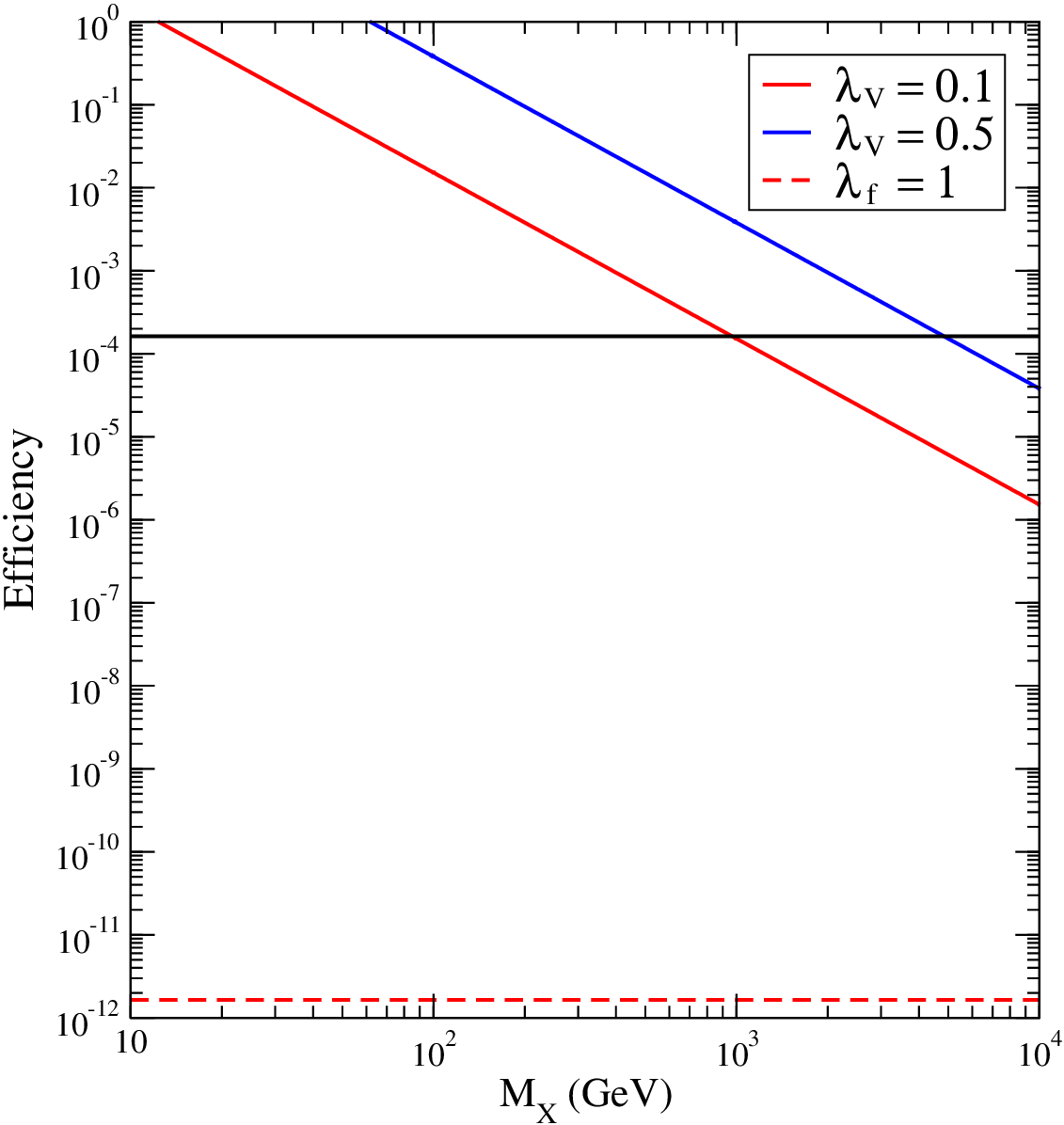}
\includegraphics[width=0.49\textwidth, clip ]{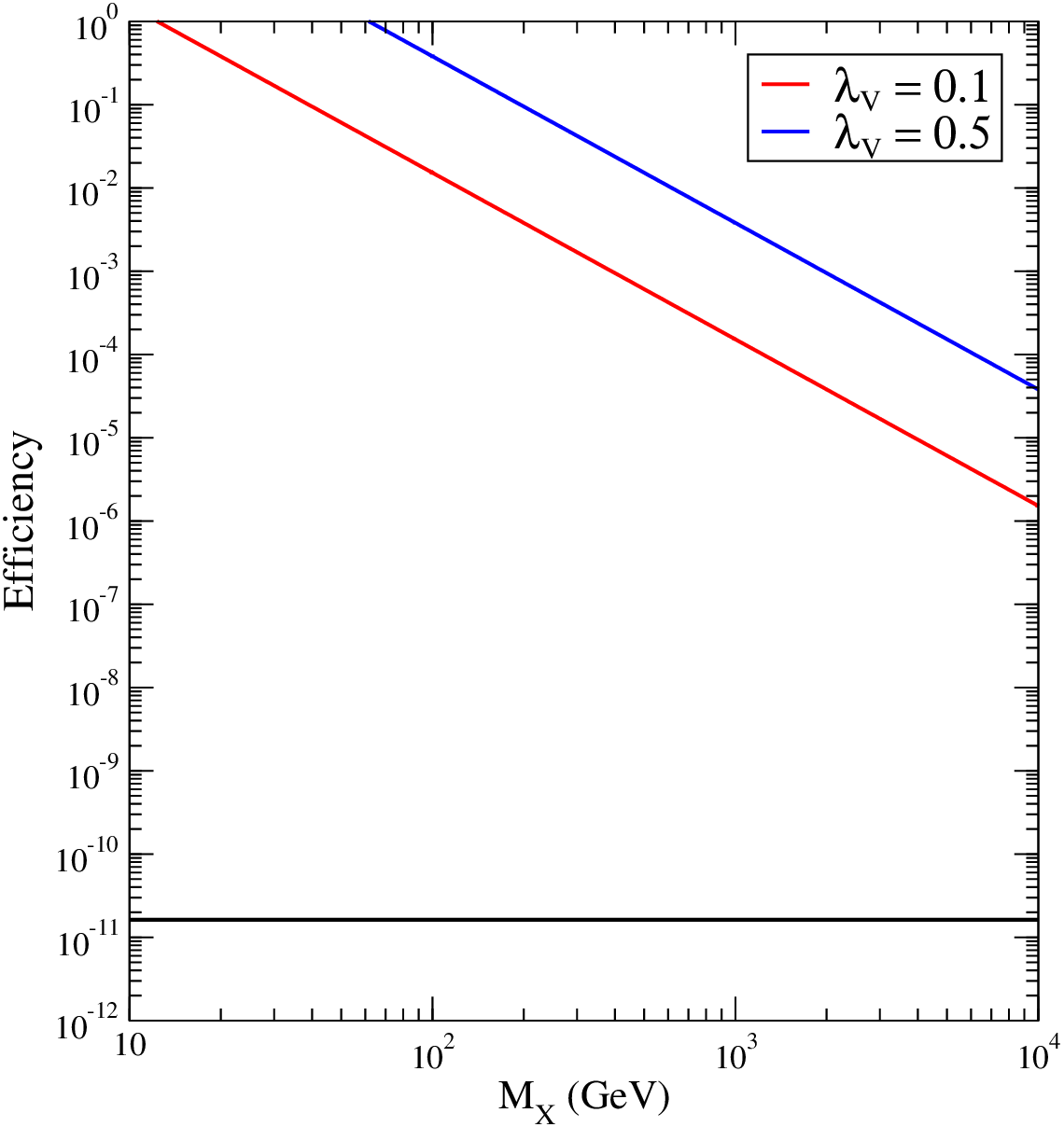}
\caption{\small 
Efficiency of vector boson (solid lines) and fermion (dashed line) particle production (scalars are too 
inefficiently produced to be shown) for a perfectly elastic collision, normalized to the most efficiently 
produced particles (in this case $W_{\mu}$ and $Z_{\mu}$) and to the energy contained in the bubble 
walls, for $\gamma_w = 10^8$ (LEFT) and $\gamma_w = 10^{15}$ (RIGHT), 
$l_w = 15/T_{\mathrm{EW}}$ and $T_{\mathrm{EW}} = 100$ GeV. The solid-black 
line corresponds to the observed dark matter comoving energy density (normalized to the energy contained 
in the bubble walls).}
\label{Fig:12}
\end{center}
\end{figure}

Since incorporating backreaction into the present analysis of particle production is extremely difficult 
and lies beyond the scope of this paper, we simply note that the relative efficiency in particle 
production for the different species in the present analysis should be roughly correct even when backreaction 
is important. Then, an estimate of the particle production in cases where some of the species 
are very efficiently produced may be obtained just by normalizing the production to the total energy in 
the bubble walls. For very elastic bubble collisions, it has been shown in section \ref{section233} 
that production of $W_{\mu}$ and $Z_{\mu}$ gauge bosons is extremely efficient, which will then leave 
very little energy left in the bubble walls for producing other particle species. The relative efficiencies 
(defined as ratios of energy in produced particles) of the different species for a perfectly elastic collision, 
normalized to the energy contained in the 
bubble walls (assuming that most of the available energy goes into producing $W_{\mu}$ and $Z_{\mu}$) 
is shown in Figure \ref{Fig:12}. A good estimate of the non-thermally generated comoving 
energy density (per particle species $\alpha$) in this case may then be given by

\be 
\label{ComEnergydensityRelative}
\Upsilon_{\alpha} \simeq \frac{20}{\sqrt{\pi\, g_*}} \, 
\frac{1}{M_{\mathrm{Pl}}\, T_{\mathrm{EW}}} \, \frac{\beta}{H} \, 
\left.\frac{\mathcal{E}}{A}\right|_{\alpha}  \, \left(\left.\frac{\mathcal{E}}{A}\right|_{W_{\mu}} \right)^{-1}
\,\frac{E_w}{A}
\ee 

The fact that this is a reliable estimate of the particle production efficiency for the case of very elastic 
collisions is due to the high-$p^2$ modes of the bubble wall carrying almost all the energy of the bubble wall.
The energy carried by the high-$p^2$ modes will then mostly go into vector boson production (their 
production efficiency at high $p^2$ is much larger than fermionic or scalar ones), result that holds 
even without incorporating backreaction into the analysis.

On the other hand, for very inelastic collisions the results from the previous section show that 
particle production is only effective for light particles ($M_X \lesssim m_h / 2$). Therefore, 
production of $W_{\mu}$ and $Z_{\mu}$ will be very suppressed in this case, along with any other 
heavy particle, and most of the available energy will go into production of SM fermions (mainly bottom
quarks) and (possibly) new light scalars or fermions with sizable couplings to the Higgs. 
 
\section{Non-thermal Multi-TeV WIMP Dark Matter} 
\label{BabyWIMP}
 
In this section we focus on the case of relatively heavy dark matter, $M_X \gtrsim$~TeV, 
and explore the conditions under which the amount of 
non-thermally produced heavy dark matter can end-up accounting for a sizable part of the observed dark 
matter relic density (dark matter may nevertheless still have a thermal component coming 
from the usual freeze-out process). 
The first condition is clearly that bubble collisions have to be fairly elastic: it has been 
shown in sections \ref{section23} that for very inelastic bubble collisions only light 
($M_X \lesssim m_h / 2$) particles are efficiently produced, while heavy particle production is extremely 
suppressed. Since fast thermalization of light species after the EW phase transition seems unavoidable\footnote{%
Dark matter may be coupled to the Higgs weakly enough as to avoid thermalization, however in that case we find it is not produced in sufficient quantities to make up for the observed relic abundance. For a discussion of the asymmetric dark matter case,  see apppendix \ref{section4}.}, for very inelastic bubble collisions either dark matter is 
not efficiently produced or it thermalizes immediately after the end of the EW phase transition, not 
having any influence on the subsequent evolution of the Universe.

For very elastic bubble collisions, the analysis from sections \ref{section23} and \ref{section24} shows 
that electroweak gauge bosons $W_{\mu}$ and $Z_{\mu}$ are most efficiently produced, and the 
relative production efficiency of heavy fermions and scalars is too low (for them to be able to 
account for a sizable part of the observed dark matter relic abundance, see Figure \ref{Fig:12}). 
This leaves heavy vector bosons with a direct coupling to the Higgs field 
as the only viable candidate for non-thermally produced dark matter during the 
EW phase transition. 

In the following we perform an analysis of heavy vector boson dark matter coupled to the Higgs, 
including an overview of thermal freeze-out and direct detection constrains from XENON100 \cite{XENON}
(see \cite{Lebedev:2011iq,Djouadi:2011aa} for more details), and a comparison between
the amount of non-thermally produced dark matter and the amount of dark matter produced 
through thermal freeze-out. We also study the evolution of the non-thermally produced dark 
matter component after the EW phase transition.
  
\subsection{Higgs-Vector Dark Matter Interplay}

Consider a vector boson dark matter candidate with mass $M_V$ and a tree-level coupling 
to the Higgs \cite{Lebedev:2011iq,Hambye:2009fg}, 

\be
\mathcal{L}_{V} = \frac{1}{2} M_V^2 \, V_\mu V_\mu + \lambda_V v_T h V_\mu V_\mu
\ee 

This coupling mediates the dark matter annihilation into Standard Model particles, as well as the 
elastic scattering on nucleons relevant for dark matter direct detection. Concerning the former process,
the Higgs boson can mediate annihilation of dark matter into electroweak gauge bosons (for heavy dark matter
they are the most important annihilation channel) through the couplings 

\be
\frac{h}{v_T} \left (2 M_W^2   W_\mu^+ W_\mu^-  + M_Z^2 Z_\mu Z_\mu \right ) 
\ee

The spin-averaged amplitude squared for the annihilation 
process $V_\mu V_\mu \to W_\mu^+ W_\mu^-$ in the limit $s \gg m_h^2$ is given by 

\be
\label{annVVWW}
|\overline{\mathcal{M}}_{VV\to W/Z,W/Z}|^2 \approx \frac{2}{3} \, \lambda_V^2 
\left (\frac{s^2}{4 M_V^4} - \frac{s}{M_V^2} + 3 \right )
\ee 

Given (\ref{annVVWW}), the thermally averaged annihilation cross section is given by

\be
\label{annVVWW2}
\left\langle \sigma v \right\rangle_{VV\to W/Z, W/Z} = \frac{z \, \lambda_V^2}{192\, \pi\, M_V^2 \, K_2(z)^2}
\int_{4}^\infty d x \, \sqrt{x - 4 }\, K_1(\sqrt{x}\, z) \, \left(\frac{x (x-4)}{4} + 3 \right)
\ee 

\noindent where $z = M_V/T$,  and $K_1(z), K_2(z)$ are Bessel functions. For $z \gg 1$, (\ref{annVVWW2}) 
reduces to 

\be
\label{annVVWW3}
\left\langle \sigma v \right\rangle_{VV\to W/Z, W/Z} \approx \frac{\lambda_V^2}{16 \,\pi\, M_V^2} 
\ee 

The thermal cross section giving rise to the observed value of the relic density 
$\left\langle \sigma v \right\rangle_{\mathrm{WMAP}} \approx 2.6 \cdot 10^{-9}\, \mathrm{GeV}^{-2}$ 
corresponds, for heavy dark matter $M_V \gg m_h$ and using (\ref{annVVWW3}), to 

\be
\label{annVVWW4}
\left[ \frac{\lambda_V}{M_V(\mathrm{TeV})}\right]_{\mathrm{WMAP}} \approx 0.3
\ee 

\vspace{3mm}

Turning now to dark matter direct detection, the spin-averaged amplitude squared for Higgs-mediated
dark matter elastic scattering on nucleons reads

\be
|\overline{\mathcal{M}}_{V N \to V N} |^2 = 
\frac{8 \,\lambda_V^2\, f_N^2\, m_N^2}{3\, (t - m_h^2)^2}
\left( 2 + \frac{(M_V^2 - \frac{t}{2})}{M_V^2} \right ) \left (2 m_N^2 - \frac{t}{2} \right) 
\approx \frac{16\, \lambda_V^2\, f_N^2\, m_N^4}{m_h^4}
\ee

\noindent
Here, $m_N \approx 0.939$ GeV is the proton/neutron mass 
and $f_N$ is the effective  Yukawa coupling of the Higgs to nucleons which, following \cite{Djouadi:2011aa},  
we take $f_N = 0.326$  based on the lattice estimate in \cite{Young:2009zb}.  
In the last step we have taken the limit $t \ll m_N^2,m_h^2,M_V^2$. 
The elastic scattering cross section then reads 

\be
\sigma_{V N \to V N} \approx \frac{\lambda_V^2 \,f_N^2 \, m_N^4}{\pi\, M_V^2\, m_h^4} 
\approx 4.2 \cdot 10^{-44} {\rm cm}^2 \,\left[ \frac{\lambda_V}{M_V(\mathrm{TeV})} \right]^2  
\ee

On the other hand, the XENON100 bound on the dark matter elastic scattering cross 
section on nucleons, for $M_V \gtrsim$ TeV is approximately

\be
\label{Xenon}
\sigma_{V N \to V N} <  M_V \cdot 2.2 \cdot 10^{-44} {\mathrm{cm}}^2
\ee 

\begin{figure}[ht]
\begin{center}
\includegraphics[width=0.65\textwidth, clip ]{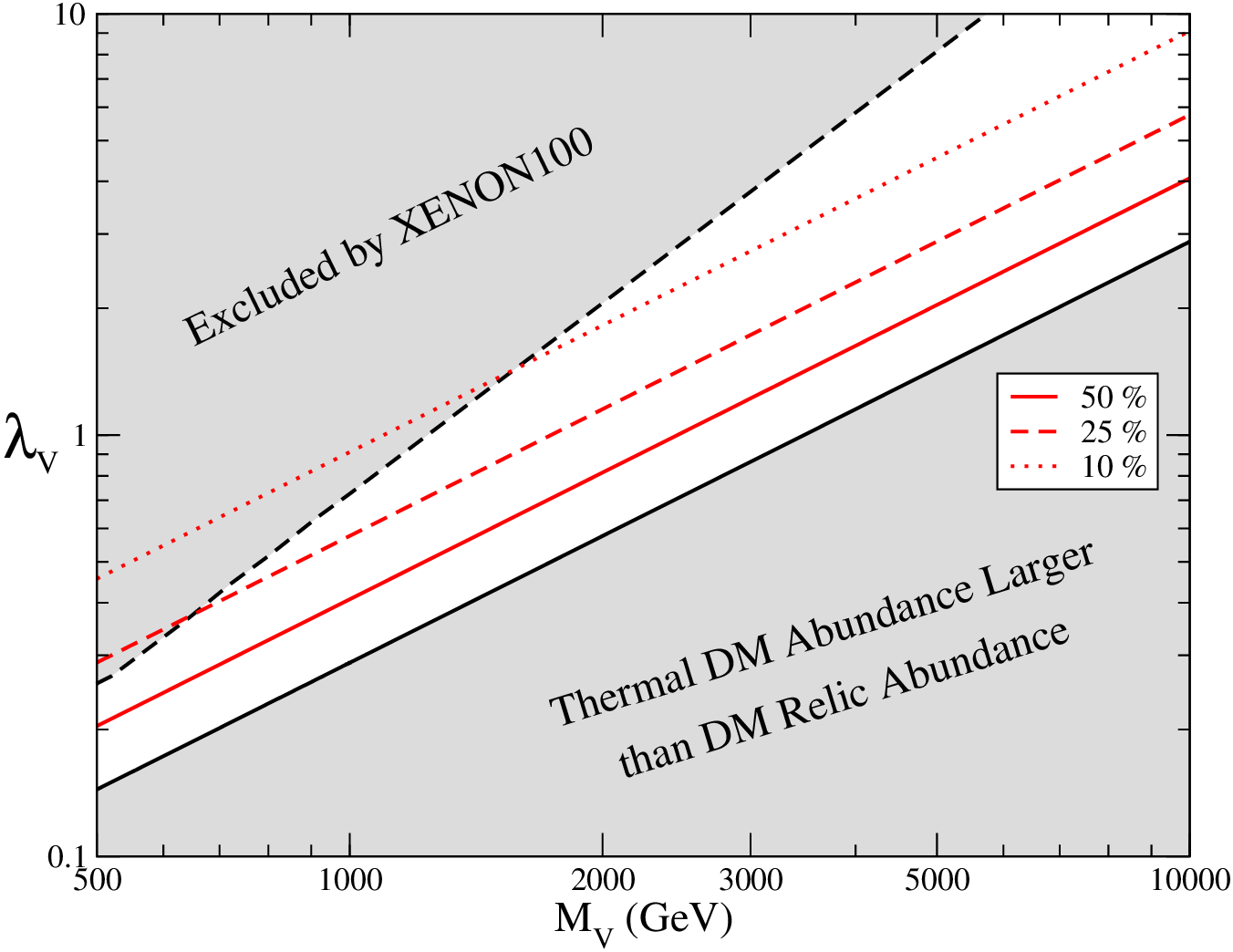}
\caption{\small 
The dashed-black line corresponds to the limits on $\lambda_V$ from XENON100 (\ref{Xenon}).
The solid-black line corresponds to the value of $\lambda_V$ for which the observed DM relic density is obtained
via thermal freeze-out (\ref{annVVWW4}): below it the thermal DM density is larger 
than the observed DM relic density (and thus this region is excluded). Above, the thermal DM density 
is only a fraction of the observed DM relic density, and the red lines show the percentage of relic density
accounted for by the thermal density.}
\label{Fig:13}
\end{center}
\end{figure}

Therefore, (\ref{annVVWW4}) and (\ref{Xenon}) leave a sizable window in the parameter space 
$(M_V,\,\lambda_V)$ for which the dark matter abundance obtained via thermal freeze-out is significantly
smaller  
than the observed dark matter relic density, and still the value of $\lambda_V$ (as a function of
$M_V$) is below the XENON100 bound (as shown in Figure \ref{Fig:13}).

\subsection{Fate of Non-Thermally Produced Vector Dark Matter}
\label{section31}

\begin{figure}[ht]
\begin{center}
\includegraphics[width=0.65\textwidth, clip ]{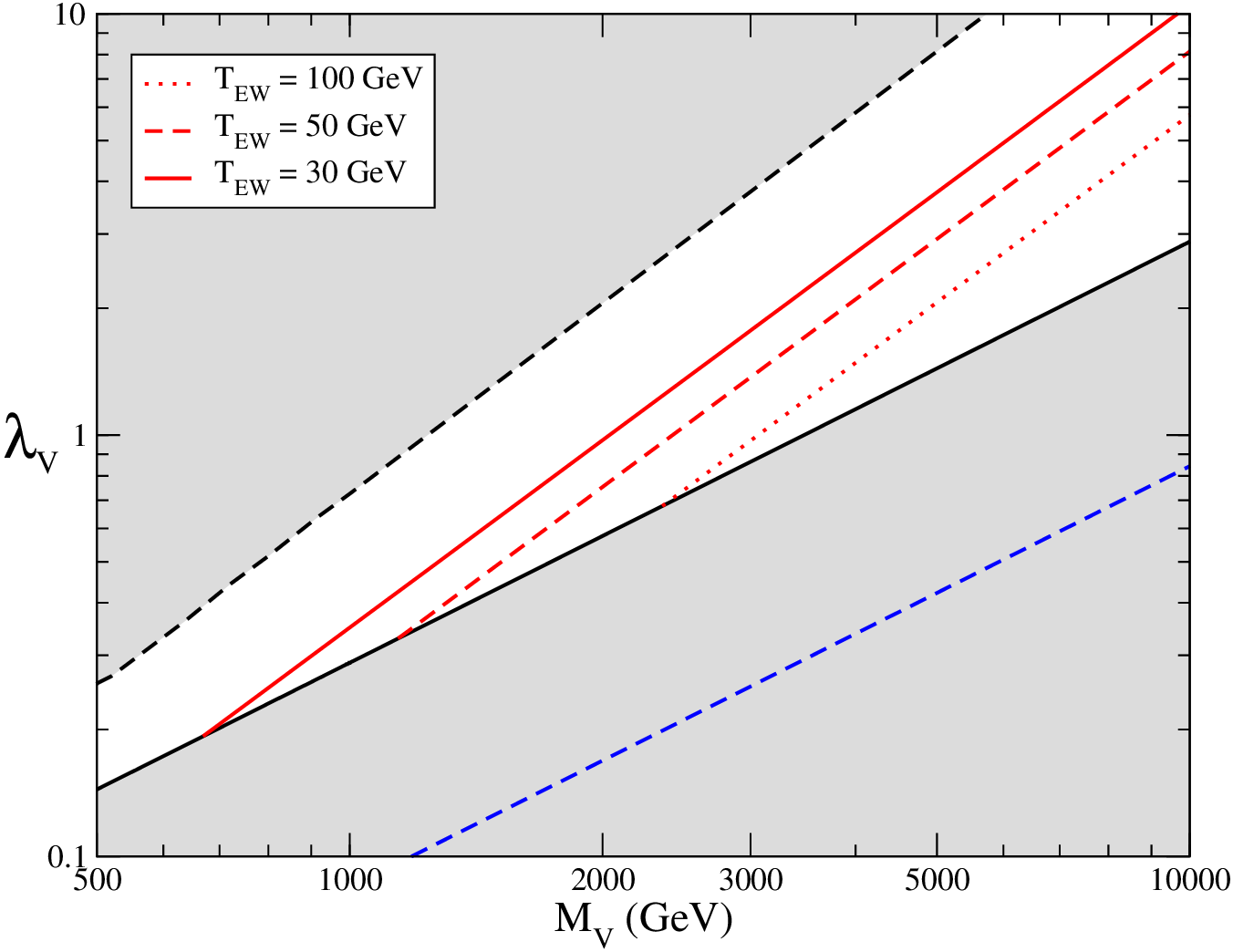}
\caption{\small 
Black lines are the same as in Figure \ref{Fig:13}. 
The dashed-blue line corresponds to the value of $\lambda_V$ needed for the non-thermally produced energy 
density in vector bosons (with a direct coupling to the Higgs) to be equal to the DM relic density, for 
$\gamma_w = 10^8$. The red lines show the values of $\lambda_V$ yielding the "non-thermal" cross section
(\ref{Boltzmann4}) (for which the final dark matter abundance, taking into account its evolution 
after non-thermal production, corresponds to the observed dark matter relic density) 
for several values of $T_{\mathrm{EW}}$.}
\label{Fig:14}
\end{center}
\end{figure}

Given the results from the previous section (summarized in Figure \ref{Fig:13}), it is fair 
to ask if, in the region of $(M_V,\,\lambda_V)$ parameter space in which the thermal component 
is not enough to account for the observed dark matter relic density, dark matter produced 
non-thermally at the EW phase transition could account for the extra needed amount. 
Using the results from production efficiency of heavy vector boson dark matter obtained in sections
\ref{section233} and \ref{section24}, we show in Figure \ref{Fig:14} the value of $\lambda_V$ (as a 
function of $M_V$) for which the amount of non-thermal vector boson production equals the observed 
dark matter relic density (dashed-blue line). Then, for values of $\lambda_V$ above the thermal 
cross section giving the observed relic density $\left\langle \sigma\, v\right\rangle_{\mathrm{WMAP}}$,
non-thermal production of heavy vector bosons is so efficient as to generate amounts of dark matter much 
larger than the observed dark matter relic density. 

Assuming that at the time of the EW phase transition vector boson dark matter is already frozen-out 
($T_{\mathrm{fo}} \simeq M_V/20 > T_{\mathrm{EW}}$), we can study the evolution of the non-thermally 
generated dark matter abundance via a simple Boltzmann equation in which the comoving dark matter number 
density $Y$ fulfills $Y(z) \gg Y_{EQ}(z)$ (with $Y_{EQ}(z)$ being the equilibrium comoving number density),
yielding

\be
\label{Boltzmann1}
\frac{d Y}{dz} = -
\alpha \, \frac{ \left\langle \sigma \, v \right\rangle  M_{\mathrm{Pl}} \,M_V}{z^2} \, 
Y(z)^2 \, \longrightarrow 
\, \frac{d y}{dz} = - \frac{1}{z^2}\, y^2(z)
\ee

\noindent with $\alpha = (4 \pi^2 \sqrt{\xi \, g_{*}})/45 \simeq 2.642$ ($g_* \sim 100$ being the 
number of relativistic degrees of freedom in the thermal plasma and $\xi \equiv 90/(32 \,\pi^3)$),
$M_{\mathrm{Pl}} = 1.2 \times 10^{19}$ GeV and $y(z) = \alpha \, \left\langle \sigma \, v 
\right\rangle  M_{\mathrm{Pl}} \,M_V \, Y(z)$. 
Integration of (\ref{Boltzmann1}) for $z > z_{\mathrm{EW}}$ yields

\be
\label{Boltzmann2}
\frac{1}{y(z)} - \frac{1}{y(z_{\mathrm{EW}})} = \frac{1}{z_{\mathrm{EW}}}- 
\frac{1}{z}\, \longrightarrow \, 
\frac{1}{y(\infty)} = \frac{1}{z_{\mathrm{EW}}} + \frac{1}{y(z_{\mathrm{EW}})}
\ee

Then, given the fact that non-thermal vector boson dark matter production is much larger 
than the observed relic density in the $(M_V,\,\lambda_V)$ region of interest, we can take the limit 
$y(\infty) \ll y(z_{\mathrm{EW}})$, obtaining

\be
\label{Boltzmann3}
y(\infty) \simeq z_{\mathrm{EW}}
\ee

From (\ref{Boltzmann3}), we immediately obtain that the value of the annihilation 
cross section that will yield the observed dark matter relic density once the non-thermally 
generated dark matter evolves after the EW phase transition is simply given by 

\be
\label{Boltzmann4}
\left\langle \sigma\, v\right\rangle = \left\langle \sigma\, v\right\rangle_{\mathrm{WMAP}} 
\frac{T_{\mathrm{fo}}}{T_{\mathrm{EW}}}
\ee

The red lines in Figure \ref{Fig:14} show the values of $\lambda_V$ yielding the correct "non-thermal" 
annihilation cross section (\ref{Boltzmann4}) for several values of $T_{\mathrm{EW}}$. 

This analysis shows that non-thermal production of multi-TeV vector boson dark matter at the EW phase transition 
(in $(M_V,\,\lambda_V)$ parameter space in which the amount of dark matter yielded by thermal 
freeze-out is not enough to account for the observed dark matter relic density) is efficient
as to generate a dark matter amount much larger than the observed relic density. This results in a reactivation 
of thermalization processes that lead to partial wash-out of the non-thermally 
generated dark matter (wash-out is not complete due to the reactivation happening 
for $T < T_{\mathrm{EW}} < T_{\mathrm{fo}}$), 
meaning that multi-TeV dark matter may have a thermal spectrum despite a 
large fraction of it having been produced non-thermally at the EW phase transition.    
As shown in Figure \ref{Fig:14}, in the presence of these non-thermally produced WIMPs, 
the relation between mass and coupling giving rise to the 
observed dark matter relic density gets modified with respect to the usual 
thermal freeze-out scenario, leading to better detection prospects in the multi-TeV region 
for future dark matter direct detection experiments. 

\section{\textit{Baby-zillas}: Super-Heavy Dark Matter from the EW Phase Transition}
\label{BabyWIMP2}

In this section we study the production of super-heavy dark matter with a mass $M_X$ 
satisfying   $M_{\mathrm{GUT}} \gg M_X \gg v_{\mathrm{EW}}$ in the bubble collisions at the end of a very strong EW phase 
transition. We call these dark matter particles {\em baby-zillas} because of many similarities  (but smaller mass) to the WIMP-zilla scenario.

\begin{figure}[ht]
\begin{center}
\includegraphics[width=0.55\textwidth, clip ]{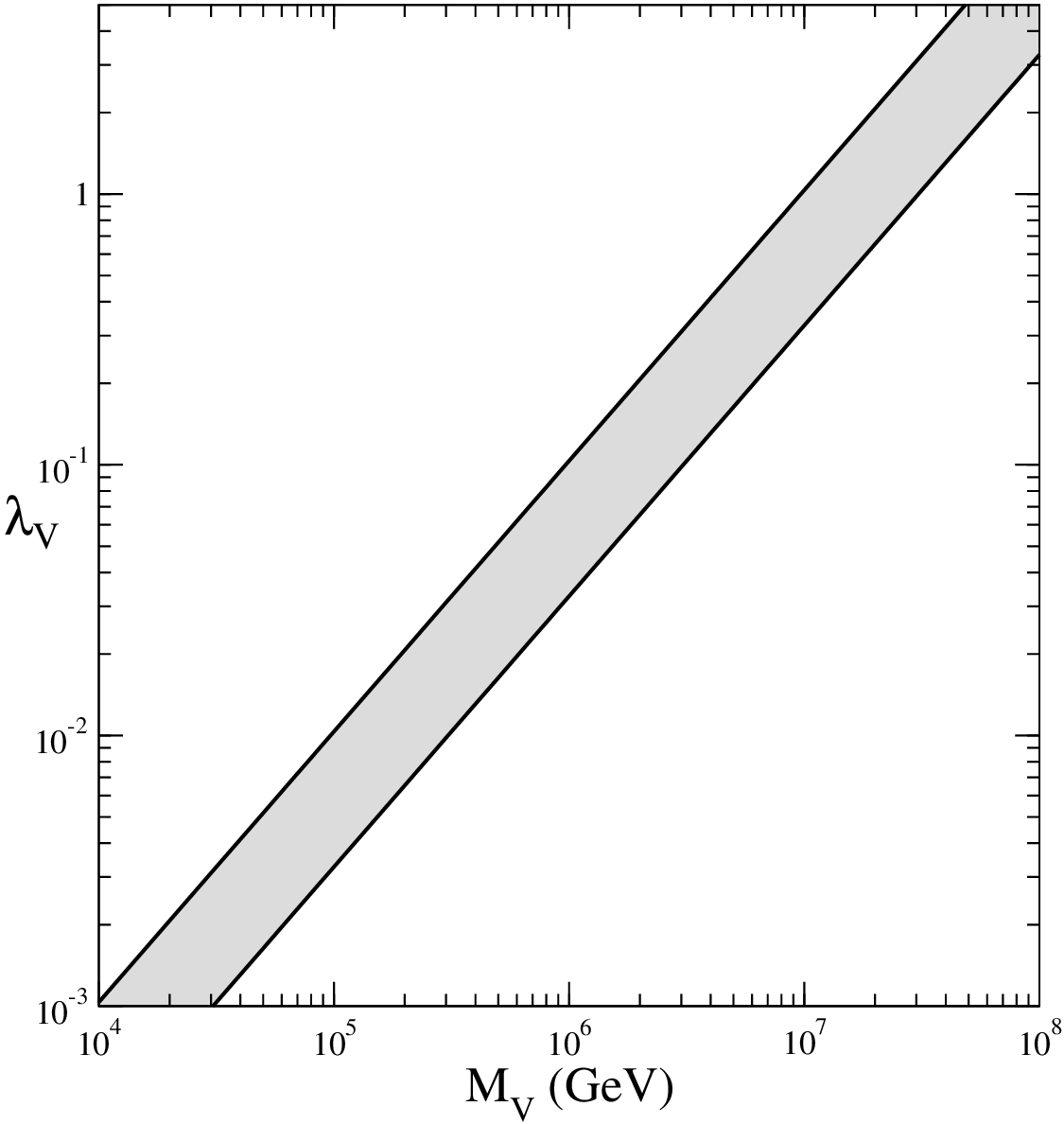}
\caption{\small 
Region in the ($M_V, \lambda_V$) parameter space for which non-thermal $V_{\mu}$ production yields the observed dark 
matter relic density for $l_w = 15/T_{\mathrm{EW}}$ (with $T_{\mathrm{EW}} = 100$ GeV) and 
$\gamma_w = 10^{14} - 10^{15}$.}
\label{Fig:15}
\end{center}
\end{figure}

From Figure \ref{Fig:12}, it can be inferred that for $\gamma_w \sim 10^{14} - 10^{15}$ non-thermal heavy vector 
boson production in elastic bubble collisions can be so efficient as to generate the observed dark matter relic 
density even for
very large dark matter masses $M_V \sim 10^6 - 10^8$ GeV and perturbative values of the coupling $\lambda_V$.  
Using (\ref{ComEnergydensityRelative}), we plot in Figure \ref{Fig:15} the region in parameter
space ($M_V, \lambda_V$) for which non-thermal $V_{\mu}$ production directly yields the observed dark matter 
relic density.

\subsection{Bounds on the Reheating Temperature After Inflation}
\label{section3}

\begin{figure}[ht]
\begin{center}
\includegraphics[width=0.55\textwidth, clip ]{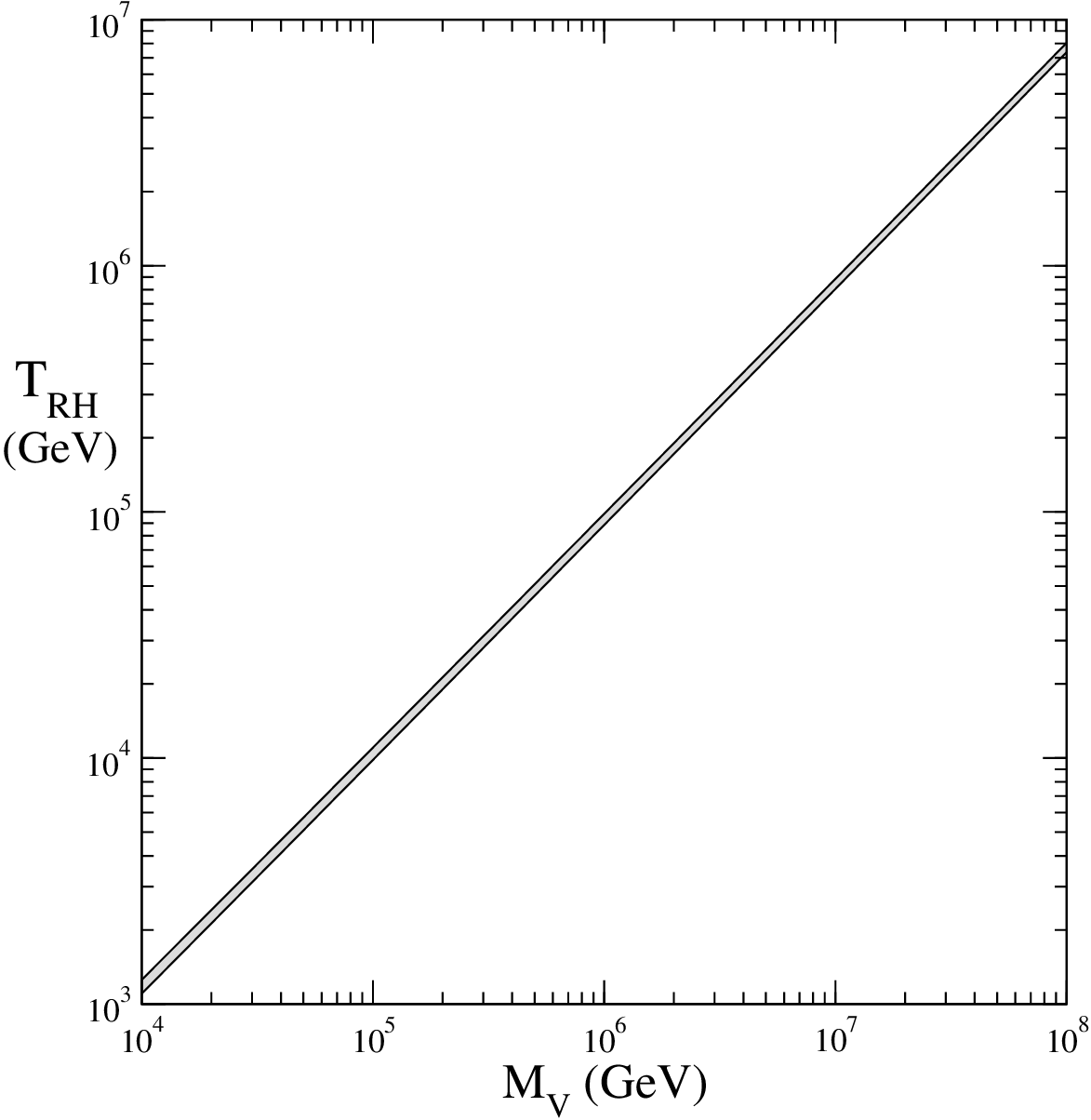}
\caption{\small 
Bounds on the Reheating temperature after inflation for the requirement that dark matter 
never reaches thermal equilibrium after inflation, namely $T_{\mathrm{RH}} \leq T_{\mathrm{fo}}$, as 
a function of the dark matter mass $M_V$, and assuming $\lambda_V(M_V)$ for which non-thermal production 
yields the observed relic abundance (as shown in Figure \ref{Fig:15}).}
\label{Fig:16}
\end{center}
\end{figure}

 A stable particle with mass $M_V \sim 10^5 - 10^8$ GeV would yield a much larger relic abundance than the observed DM relic density.  were it in thermal equilibrium 
at some stage after inflation.  
For such a massive species, the annihilation  cross is  always smaller  than the one needed to yield the observed DM relic 
density through thermal freeze-out. It is then needed that this particle species never reached thermal 
equilibrium after the end of inflation. This sets an upper bound on the reheating temperature after inflation, 
specifically $T_{\mathrm{RH}} < T_{\mathrm{fo}}$ (with $T_{\mathrm{fo}}$ being the temperature below which 
the particle is decoupled from the thermal plasma). For a heavy vector boson $V_{\mu}$ 
annihilating into $SU(2)$ gauge bosons (the most important annihilation channel in this case) through the Higgs,
$T_{\mathrm{fo}}$ satifies

\be
\frac{M_V}{T_{\mathrm{fo}}} \simeq 20.4 + \mathrm{Log}\left( \frac{M_V}{100 \,\mathrm{GeV}}\right) +
\mathrm{Log}\left( \frac{\left\langle \sigma v \right\rangle}{10^{-9}\,\mathrm{GeV}^{-2}}\right)
\ee

\noindent where the thermally averaged annihilation cross section $\left\langle \sigma v \right\rangle$ 
is given by (\ref{annVVWW2}). In Figure \ref{Fig:16} we plot 
the minimum value of $z$ (corresponding to the maximum allowed value of the reheating 
temperature $T_{\mathrm{RH}}$) as a function of the mass $M_V$ for the range of $\lambda_V$ values giving rise 
to the observed dark matter relic abundance for $\gamma_w = 10^{14} - 10^{15}$ (see Figure \ref{Fig:15}). We see that
the upper bound on $T_{\mathrm{RH}}$ is relatively insensitive to the precise value of $\gamma_w$, and roughly scales
as $T^{\mathrm{max}}_{\mathrm{RH}} \sim M_V/10$.   

\section{Conclusions}

Dark matter may have been efficiently produced at the end of a first order EW phase transition if it has a 
large coupling to the Higgs field. In this paper we investigated the conditions for this non-thermal 
production mechanism to account for most of dark matter in the Universe. 
We considered scalar, fermion and vector dark matter coupled to the SM through the Higgs (either via a 
direct, tree-level interaction or an effective Higgs-portal coupling), and found that 
production of vector bosons directly coupled to the Higgs is most efficient, while for scalar and fermions 
most of the energy stored in the bubble walls is bound to be released into production of SM particles. 
This analysis singles out vector dark matter in the present context. 
 
For very inelastic bubble collisions only dark with $M_X \lesssim 100$ GeV can 
be efficiently produced, while production of heavier dark matter is extremely suppressed. 
Unfortunately, for a dark matter mass in this range, we did not find a way to avoid subsequent 
thermalization and the wash-out of the non-thermal component, and therefore in this case dark matter 
production at the EW phase transition is irrelevant.     
The situation is quite different for highly elastic bubble collisions. 
In that case, dark matter with  $M_X \gg 100$ GeV can be efficiently produced for the so-called 
{\em runaway} bubbles, that expand with a very large $\gamma$-factor.  

We have identified two scenarios where wash-out of dark matter produced at the EW phase transition can be naturally 
avoided. One has dark matter in the multi-TeV range, which makes it possible for 
non-thermally produced dark matter to remain out of thermal equilibrium after the EW phase 
transition. We determined the region in the parameter space of dark matter mass and coupling to the Higgs 
where the correct relic abundance is reproduced. For a given mass, the coupling has to be {\em larger} than 
in the usual thermal freeze-out scenario for Higgs portal dark matter, which can be especially relevant 
for direct detection searches, as it opens the possibility of detecting a signal from multi-TeV 
non-thermal dark matter in the near future by XENON100 and LUX experiments.  
The other scenario is {\em baby-zilla} dark matter with $M_X \sim 10^6$-$10^8$ GeV.  
Surprisingly enough, such super-heavy dark matter can be produced in important quantities at the end of a strongly 
first-order EW phase transition, provided the dark matter coupling to the Higgs is large, and the $\gamma$ factor 
of the bubble walls is near its maximal value of $\gamma_w \sim 10^{15}$.  
In order for the baby-zillas to be a viable dark matter candidate, they must have never reached 
thermal equilibrium, which then constrains the reheating temperature after inflation in this scenario.

\section*{Acknowledgments}

We specially thanks Francesco Riva for very useful discussions and collaboration
in the early stages of this work, and also Thomas Konstandin, Michel Tytgat, Yann Mambrini, Stephan
Huber and Stephen West for discussions and comments. The work of J.M.N. is supported by  
the Science Technology and Facilities
Council (STFC) under grant number ST/J000477/1.

\appendix
\renewcommand{\theequation}{\Alph{section}.\arabic{equation}} 
\setcounter{section}{0} 
\setcounter{equation}{0}

\section{Asymmetric Dark Matter Production}
\label{section4}

We now explore the possibility of asymmetric dark matter production during the EW phase 
transition, together with the viability of this mechanism as a way to avoid wash-out 
of non-thermal production for relatively light dark matter (and any other light species 
in general). 

For multi-component dark matter ($X = X_{\alpha}$), an asymmetry in the number densities of 
$X_{\alpha}$ and $\overline{X}_{\alpha}$ may be generated during the particle production. 
We will analyze in detail below the generation of this asymmetry for scalars 
($X_{\alpha} = S_{\alpha}$). Then, in section 
\ref{section42} we study the evolution of the generated asymmetries after the EW phase transition.

\subsection{Decay Asymmetries: Producing a Dark Matter Asymmetry}

Let us consider a set of $N_{i}$ real scalars $h_i$ (that includes the field(s) involved in 
the EW phase transition) coupled to a set of $N_{\alpha}$ complex scalars $S_{\alpha}$ via a 
trilinear interaction. The relevant part of the lagrangian is

\be
\label{S-lagrangian}
-\Delta \mathcal{L} = m^2_{\alpha}\, S^*_{\alpha}S_{\alpha} 
+ C_{i\alpha\beta} \,h_i \, S^*_{\alpha}S_{\beta} + V(h_i)
\ee
 
\noindent where by hermiticity $C_{i\alpha\beta} = C^*_{i\beta\alpha}$ 
(it follows that $C_{i\alpha\alpha}$ are real, but $C_{i\alpha\beta}$ with 
$\alpha \neq \beta$ can be complex), and the mass matrix for the scalars 
$S_{\alpha}$ is taken to be diagonal without loss of generality. 
We also consider a possible term $\mu_{ij}\, h_i\, h^2_j$ appearing in $V(h_i)$. 
The lagrangian (\ref{S-lagrangian}) incorporates a $Z_2$ symmetry that makes the lightest of the 
scalars $S_{\alpha}$ stable, which may then be a suitable dark matter candidate.
In order for an asymmetry in the production of $S_{\alpha}$ and $S^{*}_{\alpha}$ to be generated,
we need a nonzero value for 

\be
\left|\mathcal{M}(h_i \rightarrow S^{*}_{\alpha}\, S_{\beta}) \right|^2 - 
\left|\mathcal{M}(h_i \rightarrow S^{*}_{\beta}\, S_{\alpha}) \right|^2 
\ee

\begin{figure}[ht]
\begin{center}
\includegraphics[width=0.7\textwidth, clip ]{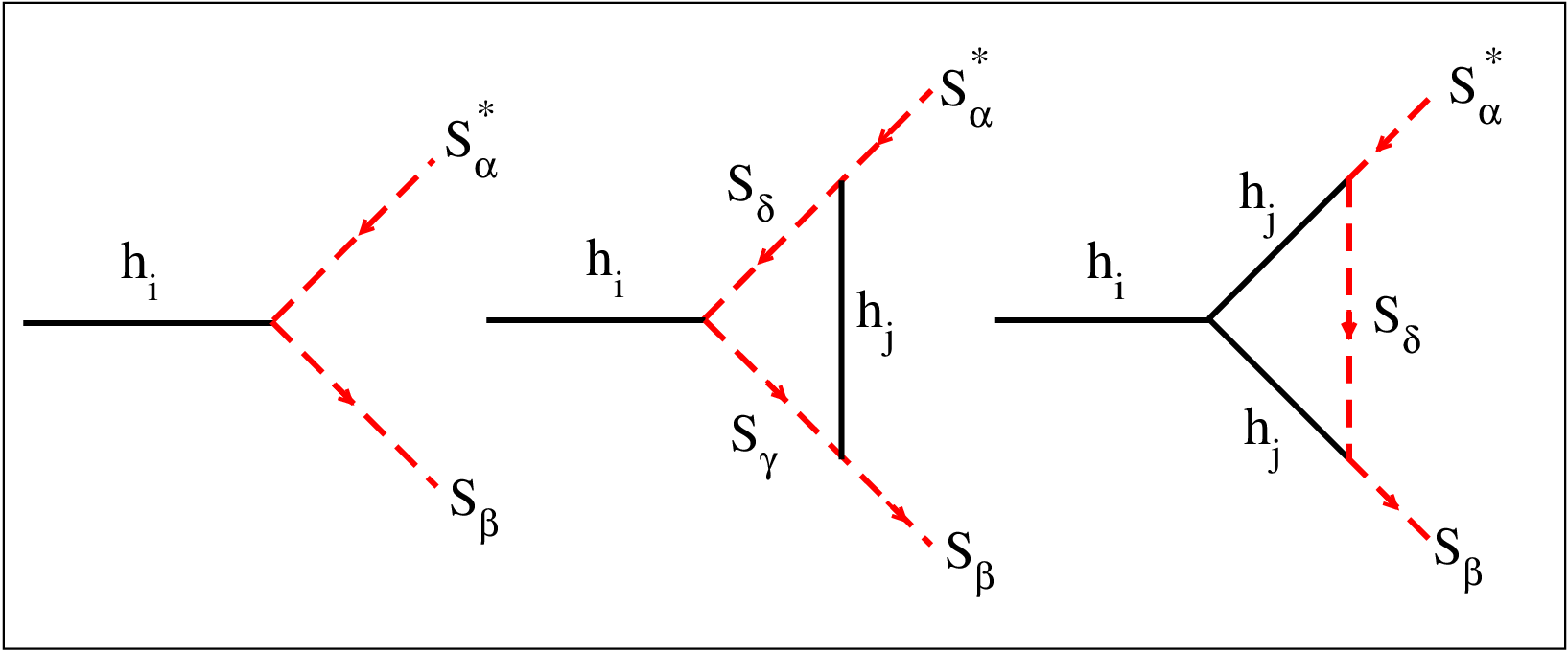}
\begin{small}
\caption{\label{figurefeynman} \small Tree-level and 1PI 1-loop contributions to 
the decay $h_i \rightarrow S^{*}_{\alpha}\, S_{\beta}$}
\end{small}
\end{center}
\end{figure}

At tree level

\be
\begin{array}{l}
\mathcal{M}^{\mathrm{Tree}}(h_i \rightarrow S^{*}_{\alpha}\, S_{\beta}) = C_{i\alpha\beta} \\
\mathcal{M}^{\mathrm{Tree}}(h_i \rightarrow S^{*}_{\beta}\, S_{\alpha}) = C^*_{i\alpha\beta}
\end{array}
\quad \Rightarrow \quad 
\left|\mathcal{M}(h_i \rightarrow S^{*}_{\alpha}\, S_{\beta}) \right|^2 = 
\left|\mathcal{M}(h_i \rightarrow S^{*}_{\beta}\, S_{\alpha}) \right|^2 
= \left|C_{i\alpha\beta} \right|^2
\ee 

\noindent and there is no asymmetry generated. At 1-loop we include the 1PI diagrams shown 
in Figure \ref{figurefeynman}. Their contribution to the 1-loop decay amplitude is

\bea
\mathcal{M}^{\mathrm{1L}}(h_i \rightarrow S^{*}_{\alpha}\, S_{\beta}) \, = \, 
-\frac{1}{16\pi^2} \sum_{j,\gamma,\delta} 
\left(C_{i\gamma\delta}C_{j\alpha\gamma}C_{j\delta\beta} \,\, I_{T} + 
 \mu_{ij} C_{j\alpha\delta}C_{j\delta\beta} \,\, \tilde{I}_{T}\right)
  \nonumber \\ 
\mathcal{M}^{\mathrm{1L}}(h_i \rightarrow S^{*}_{\beta}\, S_{\alpha})\,  = 
\,-\frac{1}{16\pi^2} \sum_{j,\gamma,\delta} 
\left(C^*_{i\gamma\delta}C^*_{j\alpha\gamma}C^*_{j\delta\beta} \,\, I_{T} + 
 \mu_{ij} C^{*}_{j\alpha\delta}C^{*}_{j\delta\beta}\,\, \tilde{I}_{T}\right)  
\eea 

\noindent where the integrals $I_{T}$ and $\tilde{I}_{T}$ correspond to

\bea
\label{Integrals}
I_{T} = \frac{-i}{\pi^2} \int d^4 k \, \frac{1}{(k^2-m_{\gamma}^2)((k+p)^2-m_{\delta}^2)((k-k_2)^2-m_{j}^2)} 
\nonumber \\
\tilde{I}_{T} = \frac{-i}{\pi^2} \int d^4 k \, \frac{1}{(k^2-m_{j}^2)((k+p)^2-m_{j}^2)((k-k_2)^2-m_{\delta}^2)}
\eea 

\noindent and can be computed in terms of the usual Passarino-Veltman 3-point scalar loop integral $C_{0}$. 
The leading order difference between $\left|\mathcal{M}(i \rightarrow \alpha^{*}\, \beta) \right|^2$ and 
$\left|\mathcal{M}(i \rightarrow \alpha\, \beta^{*}) \right|^2 $ is due to the interference between the 
tree level and 1-loop decay amplitudes. We obtain

\bea
\label{S-asymmetry}
\left|\mathcal{M}(h_i \rightarrow S^{*}_{\alpha}\, S_{\beta}) \right|^2 - 
\left|\mathcal{M}(h_i \rightarrow S^{*}_{\beta}\, S_{\alpha}) \right|^2  = 
\quad \quad \quad \quad \quad \quad \nonumber \\
\frac{1}{4\,\pi^2} \sum_{j,\gamma,\delta} 
\left\lbrace \mu_{ij} \, \mathrm{Im}\left[C^*_{i\alpha\beta}C_{j\alpha\delta}C_{j\delta\beta}\right] 
\mathrm{Im}\left[ \tilde{I}_{T} 
\right]
 + \mathrm{Im}\left[C^*_{i\alpha\beta}C_{i\gamma\delta}C_{j\alpha\gamma}C_{j\delta\beta}\right] 
\mathrm{Im}\left[ I_{T} \right] 
 \right\rbrace
\eea 

An interplay between ``weak" and ``strong" phases (complex couplings and imaginary part of a 1-loop 
integral due to particles in the loop going on-shell) is then needed to get CP violation in the decay. 
For $N_{i} = 1$ ($i = j = 1$) both terms on the right-hand side of (\ref{S-asymmetry}) will vanish 
for $N_{\alpha} < 3$ but may be zonzero for $N_{\alpha} \geq 3$. For the case of two fields 
$h_i$ ($N_{i} = 2$), already with two scalars ($N_{\alpha} \geq 2$) it is possible to
obtain an asymmetry. 

To obtain the total combined production of $S_{\alpha}$ and $S^{*}_{\alpha}$ particles, 
we will just consider the tree level contribution to $h_i \rightarrow S^{*}_{\alpha}\, S_{\alpha}$,
$h_i \rightarrow S^{*}_{\alpha}\, S_{\beta}$ and $h_i \rightarrow S^{*}_{\beta}\, S_{\alpha}$. 
We then get 

\be
\label{SymmetricProductionScalars}
\mathrm{Im}\left[\tilde{\Gamma}^{(2)} \left( \chi \right) \right]_{\alpha} = 
\left|C_{i\alpha\alpha} \right|^2 \int d\Pi_{\alpha\alpha} + 
\left|C_{i\alpha\beta} \right|^2 \int d\Pi_{\alpha\beta} 
\ee

\noindent with

\be
\label{SymmetricProductionScalars2}
\int d\Pi_{\alpha\alpha} = \frac{1}{8 \pi} \, \sqrt{1- \frac{4\,m_{\alpha}^2}{\chi}} \,
\Theta \left(\chi- 4 m_{\alpha}^2\right) 
\ee

\be
\label{SymmetricProductionScalars3}
\int d\Pi_{\alpha\beta} = \frac{1}{8 \pi} \, 
\sqrt{1-2 \frac{m_{\alpha}^2 + m_{\beta}^2}{\chi} + \frac{(m_{\alpha}^2 - m_{\beta}^2)^2}{\chi^2}}
\, \Theta \left(\chi -(m_{\alpha}+m_{\beta})^2\right) 
\ee

For the asymmetry in the production of $S_{\alpha}$ and $S^{*}_{\alpha}$ particles we obtain

\be
\label{AsymmetricProductionScalars}
\mathrm{Im}\left[\tilde{\Gamma}^{(2)} \left( \chi \right) \right]^{\mathrm{Asym}}_{\alpha} = 
\left.\left(\left|\mathcal{M}(h_i \rightarrow S^{*}_{\alpha}\, S_{\beta}) \right|^2 - 
\left|\mathcal{M}(h_i \rightarrow S^{*}_{\beta}\, S_{\alpha}) \right|^2 \right) \right|_{p^2 = \chi} 
\int d\Pi_{\alpha\beta} 
\ee

\noindent with $\left|\mathcal{M}(i \rightarrow \alpha^{*}\, \beta) \right|^2 - 
\left|\mathcal{M}(i \rightarrow \alpha \, \beta^{*}) \right|^2$ 
given by (\ref{S-asymmetry}). 

\subsection{Fate of the Generated Asymmetric Abundance}
\label{section42}

The asymmetric dark matter production process outlined in the previous section will generically 
result in asymmetries for the comoving number densities for particles and antiparticles of the 
different species $\Delta_{\alpha} \equiv Y_{X_{\alpha}} - Y_{X^*_{\alpha}} \neq 0$ at the 
end of the EW phase transition. Note however that the $Z_2$ symmetry
forces the sum of the asymmetries of the different species $X_{\alpha}$
to vanish

\be
\label{AsymSum}
\sum_{\alpha} \Delta_{\alpha} = 0
\ee    

After the EW phase transition, the comoving number densities for the different species 
$Y_{X_{\alpha}}$ will evolve according to a system of coupled Boltzmann equations. 
Denoting the symmetric and asymmetric part of the comoving number densities for particles 
and antiparticles of the different species by
$\Xi_{\alpha} \equiv Y_{X_{\alpha}} + Y_{X^*_{\alpha}}$ and $\Delta_{\alpha}$, we can write

\bea 
\label{BoltzmannSym}
z \,H(z)\, \frac{d\, \Xi_{\alpha}}{d z} & = & - \frac{s}{2} 
\left\langle \sigma \, v\right\rangle_{\alpha + \alpha^* \rightarrow \mathrm{SM}}
\left[\Xi_{\alpha}^2 - \Delta_{\alpha}^2 - \left(\Xi^{\mathrm{Eq}}_{\alpha}\right)^2 \right]\nonumber \\
& \, & - \frac{s}{2} \sum_{\beta \neq \alpha}
\left\langle \sigma \, v\right\rangle_{\alpha + \beta^* \rightarrow \mathrm{SM}}
\left[\Xi_{\alpha}\Xi_{\beta} - \Delta_{\alpha}\Delta_{\beta} - \Xi^{\mathrm{Eq}}_{\alpha} \Xi^{\mathrm{Eq}}_{\beta} \right]\nonumber\\
& \, & - \sum_{\beta \neq \alpha}
\left\langle \sigma \, v\right\rangle_{\alpha + \mathrm{SM} \rightarrow \beta + \mathrm{SM}}
\left[\Xi_{\alpha} - \frac{\Xi^{\mathrm{Eq}}_{\alpha}\, \Xi_{\beta}}{\Xi^{\mathrm{Eq}}_{\beta}}\right]\nonumber \\
& \, & - \sum_{\beta \neq \alpha}
\Gamma_{\alpha \rightarrow \beta + \mathrm{SM}}
\left[\Xi_{\alpha} - \frac{\Xi^{\mathrm{Eq}}_{\alpha}\, \Xi_{\beta}}{\Xi^{\mathrm{Eq}}_{\beta}}\right]
\eea

\bea 
\label{BoltzmannAsym}
z \,H(z)\, \frac{d\, \Delta_{\alpha}}{d z} & = & - \frac{s}{2} \sum_{\beta \neq \alpha}
\left\langle \sigma \, v\right\rangle_{\alpha + \beta^* \rightarrow \mathrm{SM}}
\left[\Delta_{\alpha}\Xi_{\beta} - \Xi_{\alpha}\Delta_{\beta} \right]\nonumber\\
& \, & - \sum_{\beta \neq \alpha}
\left\langle \sigma \, v\right\rangle_{\alpha + \mathrm{SM} \rightarrow \beta + \mathrm{SM}}
\left[\Delta_{\alpha} - \frac{\Xi^{\mathrm{Eq}}_{\alpha}\, \Delta_{\beta}}{\Xi^{\mathrm{Eq}}_{\beta}}\right]\nonumber \\
& \, & - \sum_{\beta \neq \alpha}
\Gamma_{\alpha \rightarrow \beta + \mathrm{SM}}
\left[\Delta_{\alpha} - \frac{\Xi^{\mathrm{Eq}}_{\alpha}\, \Delta_{\beta}}{\Xi^{\mathrm{Eq}}_{\beta}}\right]
\eea

where $s$ is the entropy density, $z = m_{L}/T$ ($m_L$ is the mass of the lightest species $X_{\alpha}$) and 
$H(z)$ is the Hubble parameter. From (\ref{BoltzmannSym}), if the annihilation processes 
are unsuppressed the symmetric comoving number densities for the various species $X_{\alpha}$ will be 
driven close to thermal equilibrium (the small departure from equilibrium being due to the presence of an 
asymmetry $\Delta_{\alpha}$) 

\be
\Xi_{\alpha} \rightarrow \sqrt{\Delta_{\alpha}^2 + \left(\Xi^{\mathrm{Eq}}_{\alpha}\right)^2}
\ee  
 
Ideally, in the absence of wash-out of $\Delta_{\alpha}$, these processes would delay 
freeze-out and still be active for $\Delta_{\alpha} \gg \Xi^{\mathrm{Eq}}_{\alpha}$, 
annihilating away the symmetric part of the number density and leading to $\Xi_{\alpha} 
\rightarrow \Delta_{\alpha}$. However, the various processes entering (\ref{BoltzmannAsym}) 
will tend to erase the asymmetries $\Delta_{\alpha}$. In particular, the process 
$X_{\alpha} + \mathrm{SM} \rightarrow X_{\beta} + \mathrm{SM}$ (responsible for kinetic equilibrium
among the different species $X_{\alpha}$) and the decay process $X_{\alpha} \rightarrow X_{\beta} + \mathrm{SM}$, 
will, if active, wash-out the asymmetries very rapidly, being also quite insensitive 
to the temperature of the Universe after the EW phase transition (the annihilation process 
$X_{\alpha} + X^{*}_{\beta} \rightarrow \mathrm{SM}$ also washes-out the asymmetries, but
is suppressed for low $T_{\mathrm{EW}}$). If these processes are in equilibrium after the 
EW phase transition, they will very rapidly drive 

\be
\Delta_{\alpha} \rightarrow \sum_{\beta \neq \alpha} \Delta_{\beta}
\ee  

\noindent which, together with (\ref{AsymSum}), leads to $\Delta_{\alpha} \rightarrow 0$. 
While it is possible for the decay process $X_{\alpha} \rightarrow X_{\beta} + \mathrm{SM}$ 
to be suppressed if the different species are quite degenerate, this automatically results 
in unsuppressed kinetic equilibrium. This seems to rule-out asymmetric production
as a viable mechanism to avoid complete wash-out of the amount of light dark matter non-thermally 
produced during the EW phase transition, the reason being that it is not possible (at least in 
this simple scenario) to keep all the processes that tend to erase the asymmetries $\Delta_{\alpha}$ out 
of equilibrium after the EW phase transition, while having an efficient particle 
production at the transition itself.  


\end{document}